\documentstyle[emulateapj]{article}

\lefthead{Cha \& Sembach}
\righthead{Vela Supernova Remnant}

\begin{document}

\newcommand{\degr}{$^{\circ}$}
\newcommand{\kms}{\,km\,s$^{-1}$}     
\newcommand{\vlsr}{V$_{\sc LSR}$}

\title{Spectroscopy and Time Variability of Absorption Lines in the Direction of the Vela Supernova Remnant\footnotemark}

\footnotetext{\noindent Based on observations obtained at the European Southern Observatory, La Silla, Chile}

\author{Alexandra N. Cha\altaffilmark{2} \& Kenneth R. Sembach\altaffilmark{2}}
\altaffiltext{2}{Department of Physics \& Astronomy, The Johns Hopkins University,
Baltimore, MD  21218; e-mail: {\it zan@pha.jhu.edu, sembach@pha.jhu.edu}}

\begin{abstract}
We present high resolution (R$\approx$75,000), high signal-to-noise
(S/N$\approx$100) \ion{Ca}{2} $\lambda$3933.663 and \ion{Na}{1} 
$\lambda\lambda$5889.951, 5895.924 spectra of 68 stars in the
direction of the Vela supernova remnant.  The spectra 
comprise the most complete high resolution, high S/N, optical
survey of early type stars in this region of the sky.
A subset of the sight lines has been observed at multiple epochs,
1993/1994 and 1996.  Of the thirteen stars observed twice, seven have
spectra revealing changes in the equivalent width and/or velocity structure
of lines, most of which arise from remnant gas.  Such
time variability has been reported previously for the sight lines towards
HD\,72089 and HD\,72997 by Danks \& Sembach (1995) and for HD\,72127
by Hobbs {\it et al.} (1991).  We have confirmed the ongoing time variability
of these spectra and present new evidence of variability in the
spectra of HD\,73658, HD\,74455, HD\,75309 and HD\,75821.
We have tabulated \ion{Na}{1} and
\ion{Ca}{2} absorption line information for the sight lines in our sample
to serve as a benchmark for further investigations of the dynamics and
evolution of the Vela SNR.

\end{abstract}

\keywords{line: profiles -- ISM: clouds --
ISM: individual (Vela Supernova Remnant) --
ISM: supernova remnants -- ISM: kinematics and dynamics}

\section{Introduction}

The Vela supernova remnant (SNR), located in the Southern Milky Way
at $l$\,$\approx$\,264\degr, $b$\,$\approx$\,$-$3\degr\ at a distance of
250$\pm$30 pc (Cha, Sembach, \& Danks 1999 -- Paper I), is
a bright extended X-ray source that is roughly
spherical in shape with an angular diameter of 7.3\degr\ (Aschenbach 1993).
In the
infrared, the remnant appears patchy indicating a disparity in the
concentration of dust throughout the region.  Near the center of the Vela
SNR is the Vela pulsar, a radio bright source thought to be the remains of
the star that exploded to create the SNR approximately 11,000 years ago
(Reichley {\it et al.} 1970).
In projection, several other structures are in close proximity to
the Vela SNR, including the Gum Nebula, the Vela Molecular Ridge,
SNR RX J0852.0 -- 4622, and the Vela IRAS Shell.

The combination of the large size of the Vela SNR with the presence of many
early type stars in this region of the sky has allowed the remnant gas to be
studied
through
 optical absorption line spectroscopy (Wallerstein \& Silk 1971;
Wallerstein {\it et al.} 1980; Jenkins {\it et al.} 1984;
Hobbs {\it et al.} 1991; Danks \& Sembach 1995).  New
high resolution
(R $\approx$ 75,000), high signal-to-noise (S/N $\ge$ 100) \ion{Ca}{2}
and \ion{Na}{1} spectra of over 60 stars in the direction of the
Vela SNR are presented herein.  With this data set, we measured a
distance of 250$\pm$30 pc to the
Vela SNR by noting which spectra contain moderate to high velocity
(\vlsr $\ge$ 25 \kms) lines associated with shocked SNR gas (Paper I).

Multi-epoch observations of some stars towards the Vela SNR have
also enabled a study of the temporal variability of the high
velocity remnant gas. 
Unlike the spectra of most sight lines in the sky,
the \ion{Ca}{2} and
\ion{Na}{1} absorption spectra for several sight lines
that penetrate the Vela SNR change
on timescales of a few years.
For instance, the spectra of these ions towards the
binary star HD\,72127 fluctuated throughout an eight year observing campaign
pursued by Hobbs {\it et al} (1991).
Danks \& Sembach (1995) also reported time variability towards two
stars in the direction of the remnant, HD\,72089 and HD\,72997.

The variable nature of the interstellar absorption lines in this region
of the sky provides a unique opportunity to study the ongoing
kinetic and chemical evolution of the SNR.
In this paper, we identify several sight lines exhibiting variability
and provide a set of high quality observations from which 
further absorption line studies of the remnant may be pursued.
The data acquisition and reduction methods are outlined
in \S\S2 and 3. 
A description of the region surrounding the Vela SNR is given 
in \S4. 
In \S5 the absorption line spectra for the
stars in the sample are presented. Additionally, the measured line widths and 
velocities obtained from our spectra and from previously
published studies of \ion{Ca}{2} and \ion{Na}{1} have been tabulated
to provide a broader context in which to study the remnant.
For sight lines that exhibit time variability,
multiple spectra from different observing runs are compared. A
discussion of the physical mechanisms at work in the gas
responsible for the temporal changes in the absorption components is in
\S6.  Concluding remarks are contained in \S7.

\section{Observations and Sample Information}
High-resolution optical spectra of 68 OB stars were obtained 
using the Coud\'e Echelle Spectrograph 
(CES) on the 1.4 meter Coud\'e Auxiliary Telescope (CAT) 
at the European Southern Observatory in February of 
1993, February of 1994, and January/February of 1996.  The instrumental setup
was similar for each observing run.  The
spectrograph was outfitted 
with the ``short'' camera and either an RCA CCD (ESO\#9, 1993, 1994) or
a Loral CCD (ESO\#38, 1996).
We employed two different configurations 
of the CES to observe either the \ion{Ca}{2} $\lambda$3933.663 (K) or 
\ion{Na}{1} $\lambda\lambda$5889.951, 5895.924 (D$_2$, D$_1$) 
and \ion{He}{1} $\lambda$5875.618 lines. The wavelength coverage
of the single echelle order observed in each setup was approximately
31\AA\ (\ion{Ca}{2}) or 50\AA\ (\ion{Na}{1}).  
Additional details about the CAT/CES can be found in Dekker {\it et al.} 
(1986) and Kaper \& Pasquini (1996). 

Multiple observations of 15--40 minutes in duration were made for each star.
In some cases, sight lines were observed at different epochs 
to search for 
time variability in the absorption lines associated with Vela SNR gas.
The fully reduced
data have high signal-to-noise ratios, typically S/N~$\ge$~100.  The 
spectra have a 2-pixel resolution $\lambda/\Delta\lambda$\,$\approx$\,75,000, 
or $\approx$4 \kms, as determined from the widths of the thorium-argon 
lamp calibration spectra taken each night (see \S3). 

Table~1 lists the 68 OB stars observed along 
with information about their Galactic positions (longitude and 
latitude), MK 
classifications, visual magnitudes, observed photometric colors, B--V color 
excesses, distances, heliocentric radial velocities ($v_{rad}$), and projected 
rotational velocities ($v\,sin\,i$).  Spectral types
from a variety of sources were adopted (see the endnotes for Table~1), with
preference given to MK classifications by Garrison, Hiltner, \& Schild (1977)
and Hiltner, Garrison, \& Schild (1969).  Whenever possible,
classifications based upon slit spectra were used, but in several cases it was 
necessary to rely upon objective prism spectral classifications (Houk 1978).
A colon is attached to the MK types listed in 
Table 1 determined by the latter method.
Comparison of the adopted classifications with others in the literature
shows that they are generally reliable to a class in both spectral type and 
luminosity, with correlated variations (in the sense that hotter main sequence
stars were sometimes classified as cooler, more luminous stars).  The impact of these classification differences in spectral 
type and luminosity on our derived spectroscopic distance estimates for the 
stars is reduced by these correlations.
The photometric colors in Table~1 also came from a number of sources, with 
preference given to values from Schild, Garrison, \& Hiltner (1983), 
Deutschman, Davis, \& Schild (1976), and Johnson {\it et al.} (1966). 
Variations
in these colors from one author to the next were typically 
0.02 magnitudes or less.  In a few cases, no published photometric colors 
other than those in the HD catalog exist.  A colon follows
these values in Table 1 and they are listed to a single decimal place.

The radial velocities in Table~1 are from Evans (1979), Wilson (1953), 
and Denoyelle (1987) and the projected rotational velocities are from 
Uesugi \& Fukuda (1982). Values of $v_{rad}$ and $v\,sin\,i$ in parentheses
are estimates based on the centroids and full widths at half maximum
intensity of the stellar \ion{He}{1} $\lambda$5875.618 lines in our spectra
when no previously published values could be found.
  
Two distances are listed for most stars in Table~1.  The first, d$_{sp}$, is a 
spectroscopic parallax estimate based upon the MK spectral 
classification, observed photometric colors, and the standard
reddening relation A$_V$\,=\,3.1E(B--V). The 
intrinsic OB star colors listed by Johnson (1963) and the absolute
magnitudes given by Walborn (1972, 1973) for stars earlier than B3 and
Blaauw (1963) for stars later than B3 were adopted.  
These distances 
have typical uncertainties of $\approx$25$\%$, but in cases where the MK type 
or photometry is less certain, the errors may be larger.
The second distance in Table~1, d$_{tp}$, is a trigonometric 
parallax distance based on Hipparcos data (Perryman {\it et al.} 
1997). 
  
The locations of the stars, as well as the location 
of the Vela pulsar (marked with an ``X''), are shown in Figure~1.
The various symbols indicate the
presence or absence of high velocity \ion{Ca}{2} absorption
towards each star.  In the bottom panel, Figure~1b, the locations of
stars are superimposed on an IRAS 100$\mu$m image of the 
Vela SNR which indicates the location of dust emission in this region.
The white areas indicate
strong emission, with the darker shades of gray showing where such emission
is weaker.  The patchiness of dust in the remnant is evident
in this infrared map.  Additionally, a ROSAT 0.75 keV X-ray contour 
(Snowden {\it et al.} 1995) arising from the extreme edge of the adiabatically
expanding shock that encircles the remnant is overplotted
in white, as is a circle indicating the location of RX J0852.0--46.22,
a recently discovered SNR (Aschenbach 1998).

\section{Data Reduction}

In addition to the science spectra obtained for each of the stars in our 
program, multiple bias exposures, internal quartz 
lamp flat-field exposures, and thorium-argon arc 
wavelength calibration exposures were obtained.  The exceptional 
stability of the CES resulted in reproducible calibration exposures 
throughout each night, as well as from night to night for identical grating 
positions. ``Ghosts'' from scattered light
caused by scratches on the CES echelle grating were carefully avoided.
Standard methods of bias and background subtraction and flat-fielding, 
were employed, all of 
which are described in detail by Sembach, Danks, \& Savage (1993). 

Thorium-argon spectra were used to convert pixel locations to 
wavelengths. At least 20  
emission lines per wavelength region were visually identified and 
wavelengths were assigned to their centroids 
using the Th-Ar line list by Willmarth (1987).
The centroids were established to 
an accuracy of $\pm$0.1 pixel.  A second order 
polynomial was fit to the wavelengths of the identified lines
as a function of pixel number.  
The RMS variations in the absolute wavelength solutions were
0.009\AA\ or 0.66 \kms\ 
in the blue (\ion{Ca}{2}) and 0.003\AA\ or 0.17 \kms\ in the red
(\ion{Na}{1}). 
The FWHM of the thorium lines yielded instrumental resolutions 
of 4.4 \kms\ for the 1993--1994 data and 3.9 \kms\ for the 1996 data. 

Many absorption line features appear in the \ion{Na}{1} spectral region 
that are due primarily  to water molecules within the 
atmosphere of the Earth.  To remove these telluric features, the
bright, nearby star $\delta$~Vel was observed 
(V = 1.96, A1 V, d = 23 pc; Hoffleit
\& Jaschek 1982), since it has a large $v\,sin\,i$ ($\approx$\,40 \kms) 
and minimal interstellar absorption.  The spectrum of $\delta$~Vel served
as a template for water line removal.  The 
spectrum of $\delta$~Vel was fit with a cubic spline function and normalized.  
The atmospheric absorption lines in the spectrum of $\delta$~Vel were
scaled by 
multiplying the optical depth of the telluric lines by a factor such that the 
strength of the lines from the $\delta$~Vel spectrum best matched 
the strength of the lines in each object spectrum at wavelengths of  
5883--5901\AA.  The object spectra were then divided
by the normalized, scaled template spectra to visibly null the 
Earth's atmospheric absorption features. 

The observed spectra were referenced to the Local 
Standard of Rest (LSR) frame as defined by Mihalas \& Binney 
(1981).  This convention assigns the Sun a velocity of 16.5 \kms\ 
in the direction $l$\,=\,53\degr, $b$\,=\,+25\degr.  Earth 
orbital velocity corrections were calculated to an accuracy of $<$ 0.01 \kms\
using a version of an algorithm originally developed by Gordon (1974).
The average LSR-to-heliocentric velocity conversion value for the data
set is $\Delta$V$_{LSR}$ $\approx$ $-$13.1 \kms, where V$_{LSR}$ =
V$_{helio}$ + $\Delta$V$_{LSR}$.
 
Once all of the spectra were assigned velocities,  
one spectrum for each object was identified as a template
and the remaining spectra of the
same species were interpolated onto the same velocity
grid.  The spectra of each object and 
each epoch of observation were then summed.  Since the spectra for each object 
were usually obtained consecutively, the differences in the velocity
scales were much smaller than the velocity resolution of the data.

Each spectrum was normalized by fitting a low order ($\le$3) polynomial
to continuum regions near the interstellar absorption.  Equivalent widths and
errors were calculated in the standard manner.  For each line,
Gaussian components were fit according to the prescription
outlined by Sembach {\it et al.} (1993).  In all cases we adopted the 
conservative approach of fitting the fewest number of components
necessary to produce a statistically meaningful description of the 
profiles (ie., $\chi$$^2_\nu$ $\lesssim$ 1).  Measurement errors were propagated through the
fitting process (see \S5).

\section{The Neighborhood of the Vela SNR}

To more fully understand the complexity of the Vela SNR
sight lines, the larger astronomical neighborhood of
the remnant must be considered.  
A schematic diagram containing the Vela SNR and
other prominent structures is shown in Figure 2.
Our spectroscopic study
probes stars in front of, within, and behind the Vela SNR.

Two other SNRs are near the Vela SNR, Puppis A (d = 2.2 kpc) centered at
(l = 260\degr, b = $-$2\degr)
(Dubner \& Arnal 1988; Reynoso {\it et al.} 1995),
and RX J0852.0--46.22 (d $\sim$ 200 pc) centered on 
(l = 266.3\degr, b = $-$1.2\degr)
(Aschenbach 1998; Iyudin {\it et al.} 1998).  Given the 
collection of SNRs, it follows that there must also be stellar 
associations in this general direction to provide suitable 
supernova candidates.
Indeed, the immediate vicinity boasts several such groupings
including: Vela OB1, Vela OB2, Vela R2, and
Trumpler 10. 
Vela OB1 is centered at (l = 265\degr, b = $-$1\degr) at a distance of 
$\sim$1.5 kpc (Sahu 1992), while Vela R2 is located at 
(l = 264.5\degr, b = $-$1.5\degr) and d $\sim$ 800 pc (Herbst 1975).
Using Hipparcos parallax data combined with radial velocity and
photometric information, Vel OB2 (l = 263\degr, b = $-$7\degr); 
diameter $\sim$ 10\degr) and Tr 10 (l = 263\degr, b = 0.6\degr);
diameter $\sim$ 14$'$)
have been determined to be at distances of
410$\pm$12 pc and 366$\pm$23 pc, respectively (de Zeeuw {\it et al.}, 1999).

Along the Galactic equator stretches the Vela Molecular
Ridge (VMR) (d $\sim$ 1$-$2 kpc).  The VMR
is a group of four warm giant molecular clouds that
was first identified by its strong CO emission.  Lending further
interest to the VMR is the possibility that it is associated with the 
spiral arms of the Galaxy, either as a bridge between the Local and
Carina arms, or as an extension of the Local arm towards the 
Carina arm (May, Murphy, \& Thaddeus 1988; Murphy \& May 1991).

Just beneath the equator, overlapping the southern edge of the VMR,
is the IRAS Vela Shell.  The IRAS Vela Shell is a gas and dust envelope
surrounding the Vela OB2 association, centered on one of its member stars, 
$\gamma$$^2$ Velorum at (l = 263\degr, b = $-$7\degr).  It was probably
created by the combined effects of supernovae and stellar winds
(Sahu 1992).

The largest feature in the Vela region is the Gum Nebula.
It is an extended H$\alpha$ emission source with
an angular diameter of 36\degr, centered on (l = 258\degr, b = $-$2\degr).
The distance to the center of the Gum Nebula is 800 pc -- equal
to the distance to the R association, Vela R2 (Herbst 1975; Sahu 1992).
The origin of the Gum Nebula is
undetermined, but current theories suggest that it could be
an evolved \ion{H}{2} region, an interstellar bubble produced by the stellar
winds of OB stars, a fossil Str{\"o}mgren sphere, or an extremely
old SNR.  (See Franco (1990) or Sahu (1992) for details on the
four possibilities.)

The proximity of such a multitude of structures combines to
produce a complicated montage.  Certainly some structures, such as 
certain stellar associations and the VMR, are distant -- beyond 1 kpc.
The Vela SNR, IRAS Vela Shell, Gum Nebula, and
RX J0852.0--46.22 all have estimated distances of 
$\sim$200--800 pc.  Future studies of the Vela region will hopefully
disentangle the structures, document how and if any of them
are interacting, and provide a definitive atlas of the area.

Knowing the neighboring structures towards Vela, we now turn attention
to the ISM that lies between the Sun and the remnant.
The sight lines first
penetrate nearby gas in the Local Bubble, an irregularly shaped region
characterized by low-density, hot, ionized gas  extending out to a
radius of $\sim$50--70 pc (Cox \& Reynolds 1987; Welsh {\it et al.} 1998).  
Extending beyond the Local Bubble, out
to $\sim$200 pc, \ion{Ca}{2} absorption spectroscopy reveals clouds in
the local interstellar medium (LISM) that have been detected with mean LSR
velocities of 0.9$\pm$9.4 \kms\
(Vallerga {\it et al.} 1993).  Along 
southern hemisphere sight lines
(Sco-Cen-Vel), LISM absorption components have been documented at 
velocities of $-$20 to $+$6 \kms\ (Crawford 1991; Welsh, Crifo, \& Lallement
 1998;
G\'enova {\it et al.} 1997).  \ion{Na}{1} spectra towards 11 stars with distances of
100--200 pc, projected on the sky to within 15\degr\ of the SNR, 
reveal absorption features with 
$-$1 \kms\ $\lesssim$ \vlsr $\lesssim$ +10 \kms\ 
(Cha {\it et al.}, 1999).
Beyond 200 pc, many of our sight lines 
encounter the accelerated and compressed clouds arising from
the expansion of the Vela SNR.

\section{\ion{Ca}{2} and \ion{Na}{1} Spectra Towards the Vela SNR}

The interstellar spectra of
stars in the direction of the Vela SNR range from very simple, 
with only a few components,
to complex, with many absorption features.  
The presence of many O and B stars in front of, within,
and behind the remnant allows for the determination of where along the
line of sight certain components arise.  For instance, \ion{Ca}{2}
spectral features observed at moderate to high velocities stem from
absorption by fast moving gas associated with the
Vela SNR.  Absorption components at low velocities 
are more likely caused by non-remnant foreground gas.

In Figure 3 we present the highest signal-to-noise \ion{Ca}{2} and
\ion{Na}{1} spectra acquired for each sight line in order of 
increasing HD number.  
Noted on each plot
are the name, distance of the star, and the LSR-to-heliocentric velocity 
conversion factor$\Delta$V$_{LSR}$ (see \S3).
The spectra of HD\,72089,
HD\,72997, HD\,73658, and HD\,74455, HD\,75309, and HD\,75821 are discussed 
in \S6, where evidence of time variability is presented.  
The spectra of the binary star HD\,72127
are also omitted from Figure 3 and are discussed in \S6.

In the literature, there are \ion{Ca}{2} 
and \ion{Na}{1} spectral data for some of the 
stars in our sample.  A compilation of previously published LSR velocity 
and equivalent width measurements
for both \ion{Ca}{2} and \ion{Na}{1} may be found in 
Tables~2 and 3 alongside new measurements.  
In the first column of Table~2,
the star name is listed.
The next four columns contain information pertaining to observations
found in the literature, including: the year of observation,
the central LSR velocity, the
equivalent width for each absorption line identified, a 
reference and an indication of the 
type of detector used for the observation.
In cases where the exact year of observation could not be determined, 
a range of possible dates is given.  All historical velocities 
were first converted to the heliocentric reference frame and then
corrected to the LSR defined in \S3.

Columns 6--11 contain information based upon our data;
observations from 1993 and
1994 are grouped in columns 6--8 while observations from 1996
are in columns 9--11.  Broad absorption features, classified as lines
fit with Doppler parameters of  20 \kms\ $\lesssim$ b $\lesssim$ 60 \kms,
are indicated by a superscript star ($^\star$) or filled circle ($^\bullet$)
following the equivalent width measurement.  
If the broad feature is centered at the
radial velocity of the star, a star ($^\star$) supersript is used to
denote probable stellar absorption.  Other broad absorption lines are 
flagged by a superscript filled circle ($^\bullet$) symbol and
arise due to broad or unresolved interstellar absorption.
Components grouped on the same horizontal line in the table
indicate that the absorption detected at multiple epochs 
probably arose from the same gas.  It was assumed that absorption features 
originated from the same gas if the measured velocities of the
lines were within 6 \kms\ of each other, 
and if the equivalent widths of the lines were of the same order of magnitude.
Some ambiguity arises from
the fact that our spectra resolve more components than previously
recorded in some cases.

Table~3, which contains new and historical \ion{Na}{1} absorption line
data, is organized similarly to Table~2.  In Table~3, however, two equivalent
widths are listed for each dataset: one for each line of the doublet 
(D$_2$, D$_1$).  Each central LSR velocity listed is the average velocity
of the D$_2$ and D$_1$ lines.

When comparing entries for a particular star
in Tables~2 and 3, it is important to keep in 
mind that apparent variability in the absorption 
line data may in some cases reflect different instrumental 
capabilities (including whether a photographic plate or CCD 
was used) and different data reduction techniques. The resolution of our
new data is superior to that of most previous
absorption line studies of the Vela SNR, which 
therefore may explain the presence of components (especially weak ones)
in our data set that had
not been reported previously.  The equivalent width errors for our
data are based upon the statistical uncertainties associated with 
Poisson noise fluctuations and are appropriate for the assumed
velocity model fit to the data.  They do not account for systematic
errors in continuum placement or background zero-level offsets.
Both of these effects are expected to be small ($\lesssim$2\%)
in these data.  The limits quoted in Tables 2-3 are 1$\sigma$
estimates of the statistical errors.

\section{Time Variability of Interstellar \ion{Ca}{2} and \ion{Na}{1} Lines}

The temporal variability of absorption
lines in the Vela SNR reveals the presence of
rapid (t $\sim$ few years) evolution and$/$or motion of
material within the remnant.  Variability of high velocity 
($\mid$ \vlsr\ $\mid$ $>$ 50 \kms) absorption lines is apparent in five stars,
while two stars exhibit changes at lower velocities.
The low incidence of detectable
variability in low velocity absorption lines may be a
selection effect since in most spectra
there are simply too many absorption lines
blended together at low velocity to observe slight
differences in the characteristics of individual components. 
Excluding Vela sight lines, there is only one other convincing
case of temporal variability of an optical absorption line.
Blades {\it et al.} (1997, see also Blades \& Penprase 1999) 
report variability in gas at $|$\vlsr$|$ $<$ 30 \kms\ toward HD\,28497,
(l = 208.78\degr, b = $-$37.40\degr).

In Figures~4--10, we highlight spectra of the sight 
lines towards HD\,72089, HD\,72127, HD\,72997, HD\,73658, HD\,74455,
HD\,75309, and HD\,75821, 
which have detectable changes in the column density of \ion{Ca}{2} 
and/or \ion{Na}{1} absorption lines and/or shifts in the 
central velocities of the lines.  Details pertaining to 
possible physical explanations for the variations are 
discussed below.  The measured velocities, \vlsr, column 
densities, N, equivalent widths, W$_{\lambda}$, and the LSR-to-heliocentric
velocity conversion factors, $\Delta$V$_{LSR}$, of the variable lines
are summarized in Table~4.  

\subsection{HD\,72089}
The \ion{Na}{1} D lines of HD\,72089 (Figure~4) contain a 
component at \vlsr\ $\approx$ +105 \kms\ whose equivalent width 
decreased by $\sim$60\% from 1993 to 1996.  Analysis of the \ion{Ca}{2} 
line at the same velocity shows a decrease in its equivalent width
of $\sim$17\% during the same period of time.   Since the amount of \ion{Ca}{2}
decreased more slowly than \ion{Na}{1}, the ratio of the 
column densities of \ion{Ca}{2} to \ion{Na}{1} increased by more
than twofold: in 1993 N(\ion{Ca}{2})/N(\ion{Na}{1}) = 3.4, while in 
1996  N(\ion{Ca}{2})/N(\ion{Na}{1}) = 7.2.  An increase in the number of ionizing photons in the +105 \kms\ gas could cause the observed 
metamorphosis of the \ion{Ca}{2} and \ion{Na}{1} lines given that the
ionization potentials (IP) of these two ions are 11.87 eV and 5.14 eV,
respectively.

\subsection{HD\,72127}
We obtained \ion{Ca}{2} absorption line spectra towards HD\,72127A
in 1994 and 1996, and towards HD\,72127B, a binary companion, in 1996
(Figure~5).  Comparison of the \ion{Ca}{2} spectra of HD\,72127A and B reveals
immediately that the \vlsr\ $\approx$ 0 \kms\ component of HD\,72127B has an
equivalent width twice that of HD\,72127A.  This dominant feature
in the spectrum of HD\,72127B is probably a blend of
absorption lines at \vlsr\ $\approx$ $-$10 and +1 \kms.  A higher
resolution spectrum of
HD\,72127B with these two lines resolved could reveal details about 
the small scale structure in the Vela region.  There may be
interstellar density inhomogeneities in the slightly
separated sight lines towards HD\,72127A and B, or perhaps the sight line
towards HD\,72127B penetrates an additional low velocity cloud that
HD\,72127A does not.
The two stars are 2500 AU apart (Thackeray 1974), separated
by more than enough distance for there to be variations in the
structure of the ISM through which their sight lines pass. It has been
documented that there exists small scale structure in the interstellar
medium revealed by the comparison of the stars in the binary system
$\mu$ Cru at a separation of 6600 AU (Meyer \& Blades 1996), and across
distances as small as 5--100 AU using unrelated background sources
(Frail {\it et al.} 1994).  
Interestingly, no subsequent variations in \ion{Na}{1}
line intensity have been observed toward $\mu$ Cru over a recent
21 month baseline (Lauroesch {\it et al.} 1998).

Absorption along the sight line towards HD\,72127A varies with time.
Analysis of our \ion{Ca}{2} spectra of HD\,72127A from 1994 and 1996,
reveals that the relative intensities of the components at \vlsr\ =
$-$10 \kms\ and $+$1 \kms\ have reversed, primarily
 because of the 11\% decrease in
equivalent width of the \vlsr\ = $-$10 \kms\ line.
During the same epoch, the two absorption lines at \vlsr\ = $-$28 \kms\
and $-$21 \kms\ increased in equivalent width by $\sim$20$\%$.
Because these three lines are so closely spaced in velocity, 
it is difficult
to assign a unique fit to the lines in the spectra.  
It is likely that the components at \vlsr\ = $-$28 \kms\
and $-$21 \kms\ are remnant gas, but we do not know
where the $-$10 \kms\ component is located along the sight line.
The \ion{Ca}{2} gas
concentration towards HD\,72127A has indeed changed
between 1994 and 1996, but the blended lines result in some
ambiguity in making a quantitative measurement of the variability.

The variable nature of the column
densities associated with the clouds towards HD\,72127A has been documented
by Hobbs {\it et al.} (1991) also.  Looking at eight \ion{Ca}{2} spectra that
Hobbs {\it et al.} obtained
from November 1981 through December 1988, the \vlsr\ $\approx$ $-$10 \kms\
component is clearly stronger than the \vlsr\ $\approx$ +1 \kms\
component (which they identify as a blend of two lines)
from November 1982 through November 1986, then vice versa in
December 1988.  As they
documented in the long time series of data they presented,
the strengths
of both the \vlsr\ $\approx$ $-$10 and +1 \kms\ lines change over time
as do the radial velocities of both absorption lines.  Neither our
data, nor the Hobbs {\it et al.} data contain evidence of a cyclic 
pattern to the changes.

The  Hobbs {\it et al.} \ion{Na}{1} spectra of HD\,72127A
also clearly exhibit variable column densities in the lines at 
\vlsr\ $\approx$ $-$10 and +1 \kms.  Since we obtained a spectrum
of HD\,72127A in 1994 only, we cannot confirm the presence of recent
variability of this ion in the high velocity gas of the Vela SNR.

\subsection{HD\,72997}
Figure~6 depicts the evolution of \ion{Ca}{2} and \ion{Na}{1} absorption 
lines toward HD\,72997 
at three epochs.  These plots clearly show shifts in the 
central line velocity of both species near 
\vlsr\ $\approx$ +190 \kms\, first reported by Danks \& 
Sembach (1995).  We reproduce their 1991 and 1993 spectra 
here, and add new spectra from 1996.  The line at \vlsr\ $\approx$ +190 \kms\
appears to be continuing its shift towards a 
larger positive velocity.  From 1991 to 1993, $\Delta$V = 2.7 
\kms\ for the high velocity \ion{Na}{1} data and $\Delta$V = 1.4 \kms\ for the 
\ion{Ca}{2} data.  Examining the epoch 1993 to 1996 we find $\Delta$V = 
1.1 \kms\ and $\Delta$V = 0.9 \kms\ for the \ion{Na}{1} and \ion{Ca}{2} data 
respectively, leading to total shifts of $\Delta$V = 3.8 \kms\ for the 
high velocity \ion{Na}{1} component and $\Delta$V = 2.3 \kms\ for the Ca 
II component over five years.
Such changes indicate that on average, that the acceleration of this
high velocity gas is decreasing with time.  The acceleration
does not appear to have caused a net change in the equivalent
width of either \ion{Ca}{2} or \ion{Na}{1} over the five-year
period.

\subsection{HD\,73658}
The intermediate velocity features at $\sim$$-$32 \kms\ and $-$15 \kms\
in the \ion{Ca}{2} spectrum of HD\,73658 decreased dramatically between
observations in 1993 and 1996.  The equivalent width of the $\sim$$-$32 \kms\
line decreased from 41$\pm$1 m\AA\ to 4$\pm$2 m\AA, a 90\% reduction, while
the $-$15 \kms\ line decreased by 53\%.  The \ion{Ca}{2} spectra from
both observations are plotted together and shown in the left panel of 
Figure 7.  Although the column density of intermediate negative velocity 
gas in this spectrum diminished since 1993, 
neither the high velocity 
absorption component at $\sim$$-$127 \kms\, nor the low and intermediate
positive velocity components changed during the same time period.
The gas at $\sim$$-$32 \kms\ and $-$15 \kms\ is 
probably patchy, with a dense region probed by the sight line in the
1993 observation and a less dense region penetrated in 1996.   
The \ion{Na}{1} D$_2$ spectrum of HD\,73658 is shown
in the right panel of Figure 7, as observed in both 1993 and 1996, 
yet there is no evidence
of time variability and there are no absorption features at intermediate
negative velocities.

\subsection{HD\,74455}
The high velocity feature at V$_{LSR}$ = $-$173 \kms\ in the \ion{Ca}{2}
spectrum of HD\,74455 decreases in equivalent width by 28\% from 1994
to 1996.  There is no change of this magnitude in any of the
other absorption components in the \ion{Ca}{2} spectrum or in the
\ion{Na}{1} spectrum.  Figure 8 illustrates that the \ion{Na}{1} spectrum
does not contain absorption at high velocities, although the 
low -- intermediate velocity profile is similar to that of \ion{Ca}{2}.
The equivalent width decrease at $-$173 \kms\ suggests that the
amount of high velocity low ionization gas along the sight line has diminished.
The sight line towards HD 74455 probes a different piece of the $-$173 \kms\
gas packet in 1996 than it did in 1993.  It does not appear that the 
reduction in column density is due to the acceleration/deceleration 
of some of the gas
since the equivalent widths of the two absorption features flanking the
$-$173 \kms\ line do not vary (see Table 2).

\subsection{HD\,75309}
In the \ion{Ca}{2} spectrum of HD\,75309 shown in Figure~9, we note three
line components that have changed between observations
in 1993 and 1996.  First, the absorption feature at 
\vlsr\ $\approx$ $-$119 \kms\ has increased in equivalent width by 25\%.
Second, in the 1993 spectrum, two absorption features are present at
\vlsr\ = +81 \kms\ and +89 \kms, with equivalent
widths of 4 m\AA\ and 7 m\AA, respectively; these
lines are not detected in the 1996 spectrum. 

The variable absorption feature at $-$119 \kms\
could result from one or both of the following mechanisms: 
the continuing compression of a patch of high velocity gas,
or the liberation of \ion{Ca}{2} due to grain destruction
at high velocities (Spitzer 1978).
We do not detect this high velocity absorption line in the \ion{Na}{1}
spectrum, nor do we see any intermediate velocity \ion{Na}{1}
absorption corresponding to the rather complicated \ion{Ca}{2} spectrum.  

We do not detect the pair of \ion{Ca}{2} absorption lines found
in the 1993 spectrum
at \vlsr\ = +81 \kms\ and +89 \kms\ in the 1996 repeat observation.
The non-detection allows us to place an upper limit on the equivalent
widths of $<$ 3 m\AA\ (3$\sigma$) in the 1996 spectrum.
With only
two observations of HD\,75309, we are not able to determine whether these
features were present for a long time and suddenly disappeared, or
whether the lines were ephemeral.  Since the 1996 spectrum
shows no trace of \ion{Ca}{2} at these velocities, the most likely
interpretation
of the presence and subsequent absence of these absorption features is
that a wisp of gas containing a column density of N(\ion{Ca}{2}) $\approx$ 
5 $\times$10$^{10}$ cm$^{-2}$ temporarily passed across the sight line.

\subsection{HD\,75821}
Our 1993 and 1996 \ion{Ca}{2} and \ion{Na}{1} spectra of HD\,75821
are displayed in Figure~10.  The plots on the right side of Figure~10
magnify a pair of spectral lines whose equivalent widths have changed
between the two observations. The central velocity of one component
is also changing.

The equivalent width of the \ion{Ca}{2} absorption line at
\vlsr\ = $-$85 \kms\ decreased by $\sim$20\% while the \ion{Na}{1}
lines associated with absorption by the same gas decreased
by $\sim$50\%.
This results in an increase in the
\ion{Ca}{2} to \ion{Na}{1} ratio of $\sim$40\%, which is likely to
occur in a region experiencing dust destruction (Spitzer 1978).  
See Danks \& Sembach (1995)
for a summary of recent work focusing on shock models
and their implications for gas phase abundances in SNR gas.

The \ion{Ca}{2} and \ion{Na}{1} absorption features
apparent near $-$99 \kms\ increased in equivalent width by $\sim$20\%
from one observation epoch to the next.  In addition, both lines
exhibit increased central velocities in 1996: for \ion{Ca}{2} 
$\Delta$V = 1.43 $\pm$0.10 \kms\ and for \ion{Na}{1} 
$\Delta$V = 1.43 $\pm$0.45 \kms.  The increase in column density of both
ions coupled with the increasing velocity of the spectral line suggests
that the high velocity gas is being accelerated.

Not only does the sight line towards HD\,75821 reveal two pockets of 
gas whose physical characteristics vary on short timescales, these 
spectra are 
unusual in another way: they serve as a counterexample to the 
``Routly-Spitzer effect''.  In 1993 (1996), the ratio 
N(\ion{Ca}{2})/N(\ion{Na}{1}) is $\sim$5 (12) times higher for the
lower velocity line at \vlsr\ $\approx$ $-$85 \kms\ than for the line at
\vlsr\ $\approx$ $-$99 \kms, while the observational trend
described by Routly \& Spitzer (1952) states that the ratio
should increase with increasing velocity.

\section{Discussion}
Multiple spectroscopic observations of a large sample of O and B stars
in the direction of the Vela SNR have enabled us to confirm and extend 
the reports of
short timescale changes in the spectra of the stars HD\,72089, HD\,72997,
and HD\,72127, as well as discover variability in the
spectra of HD\,73658, HD\,74455, HD\,75309 and HD\,75821.  
Many of the cases of variability have occured
in spectral lines at high velocities, indicating that the changes in the
absorption profiles are due to the evolving SNR.  Thirteen stars were 
observed at two epochs.  Of these,
three were known to exhibit variability.  Eliminating these stars, four of
ten randomly chosen sight lines revealed variability.  The sample is too
small to confidently state the probability of observing variability, 
but if more stars were
observed at an additional epoch, the nature of variability in the 
remnant could be explored in greater detail.  If
a large percentage of spectra changed over time, one could conclude that
1) changing conditions persist over ''long'' time periods and/or 2) changing
conditions are ubiquitous in the remnant.  On the other hand, if a
small fraction of the spectra observed changed over time, the conclusions
that could be reached include 1) changing conditions are short lived (less
than a few years) and/or
2) variability along sight lines through the Vela SNR is rare.

We note that there has been a recent discovery of an X-ray bright
supernova remnant (RX J0852.0-4622) in the direction of Vela (Aschenbach
1998).  Both the X-ray and $\gamma$-ray emissions from this object indicate
that it is within 1 kpc of the Sun, and perhaps as close as 200 pc 
(Aschenbach 1998; Iyudin {\it et al.} 1998).  If the remnant lies as close
as 200
pc, then it is possible that RX J0852.0$-$4622 and the Vela SNR are 
physically related. One of our sight lines, HD\,75821, passes very near
to RX J0852.0$-$4622.  The sight line contains high velocity gas, but there
is nothing peculiar about the optical absorption properties compared to
those of other high velocity gas features toward other Vela stars.  

An extensive study of 
\ion{Ca}{2} and \ion{Na}{1} in high velocity gas over a period of 
time may provide insight into the dynamics and evolution of 
the Vela SNR since the absorption properties of numerous sight lines
through the remnant appear to be changing over time periods of several years.
We hope that the compilation of spectroscopic
data in this paper will be useful for tracking the development 
and evolution of absorption lines traversing the SNR.  Studies of
dominant ionization stages along 
some of these same sight lines with the Hubble Space Telescope
would be useful for studying 
the changing physical conditions.  In addition, repeat observations
separated by a few months, and others spaced by $\sim$5\+ years would allow
the timescale of the changes to be determined more accurately, resulting
in a more thorough understanding of the evolution of the Vela SNR gas.

It would also be desirable to monitor the spectra of a modest sample
of stars outside the Vela region to determine if variability is common in 
the diffuse ISM.  Evidence of variability in such regions could have
a substantial impact on studies relying on comparisons of data taken
at different epochs.

\acknowledgments  
We thank Anthony Danks for his assistance with the observations.
We appreciate comments on this work from our colleagues in the 
Department of Physics and Astronomy at the Johns Hopkins University.
We acknowledge use of the Simbad Database at the Centre Donn\'ees 
astronomiques de Strasbourg (http://simbad.u-strasbg.fr/Simbad), and 
use of the NASA SkyView facility
located at the Goddard Space Flight Center (http://skyview.gsfc.nasa.gov).
KRS and ANC appreciate support from NASA Long Term Space Astrophysics Grant 
NAG5-3485 and grant GO-06413-95A from the Space Telescope Science Institute,
which is operated by AURA under NASA contract NAS5-26555.

\newpage

\begin{center}Figure Captions\end{center}

\noindent
Figure~1.  A plot of the stars in our sample a) as projected onto the sky,
and b) as projected onto an IRAS
100 $\mu$m image of the Vela region shown in logarithmic gray-scale.
The contour shown is a ROSAT 0.75 keV contour with 4 $\times$ 10$^{-4}$ counts s$^{-1}$ arcmin$^{-2}$.
Different symbols indicate the maximum velocity gas
observed along a particular sight line.  The circle in panel b) shows the 
location and size of SNR RX J0852.0--46.22.

\noindent
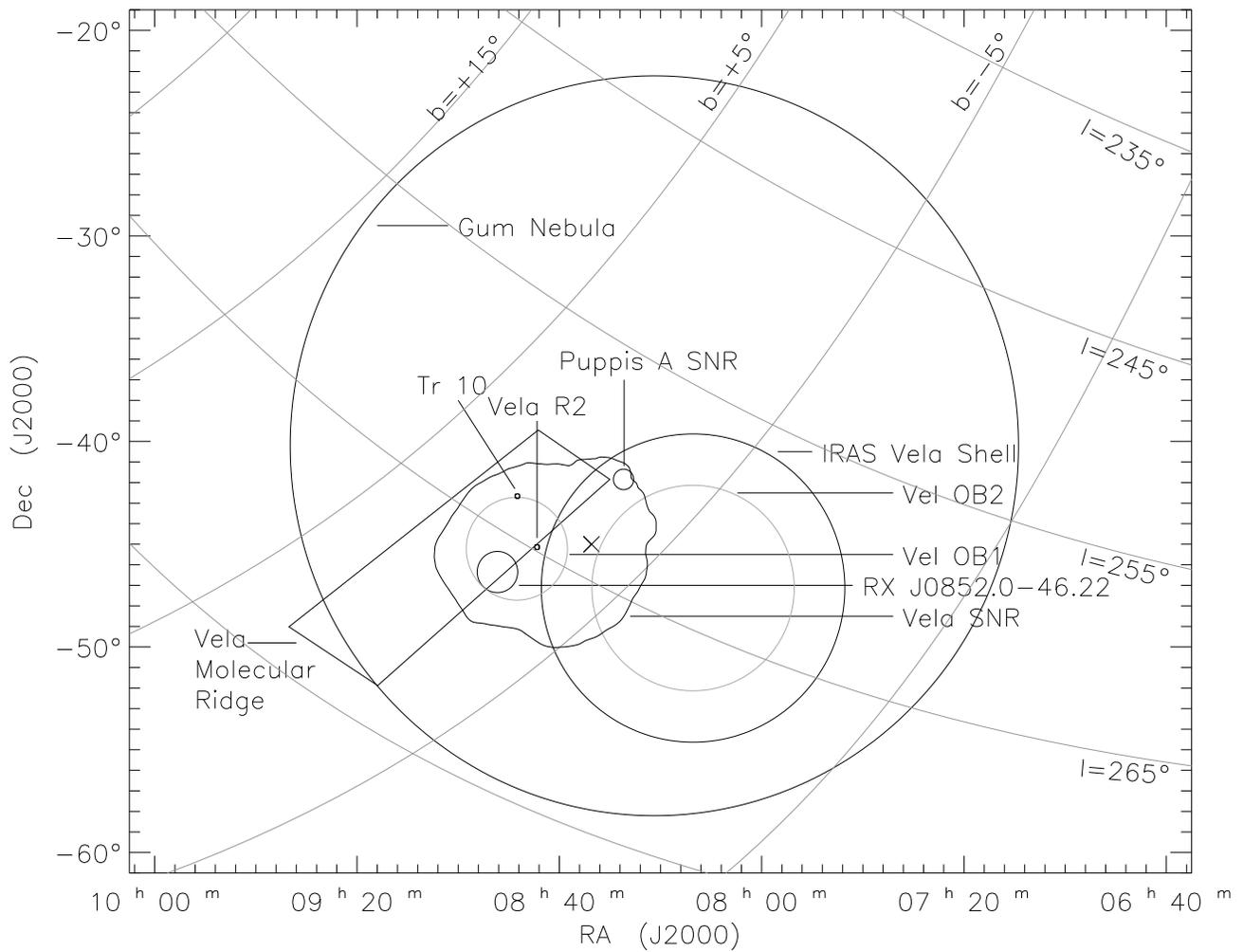
Figure~2.  A schematic diagram showing the extended structures and stellar
associations
present in the direction of the Vela SNR.  The 'X' indicates the location
of the Vela Pulsar. 

\noindent
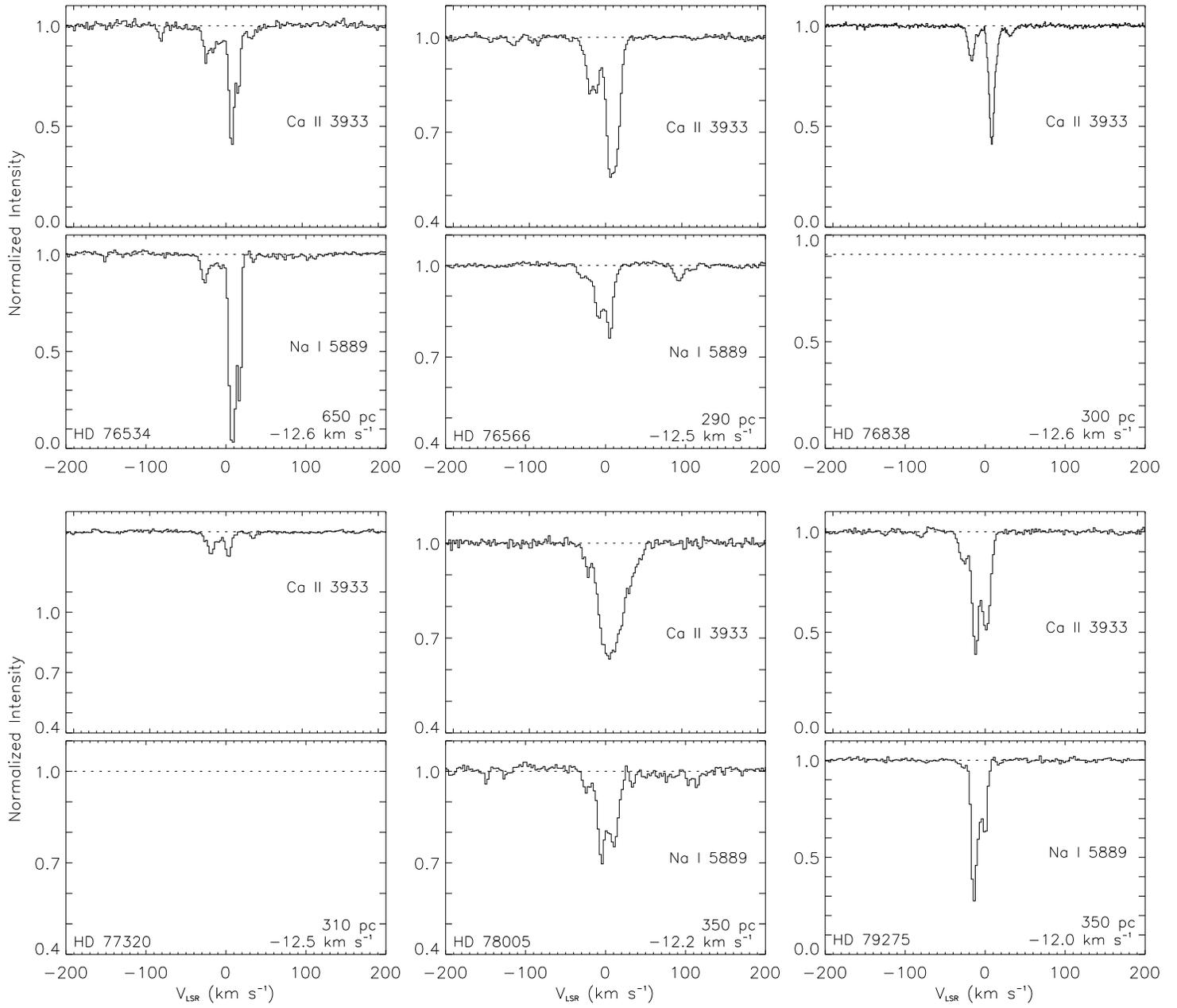
Figure~3.  Plots of the normalized \ion{Ca}{2} and \ion{Na}{1} spectra
in our sample.  If only a \ion{Ca}{2} spectrum is shown for a given
star, the corresponding \ion{Na}{1} spectrum was not obtained.
The LSR-to-heliocentric velocity
conversion factor, $\Delta$V$_{LSR}$, where
V$_{LSR}$ = V$_{helio}$ + $\Delta$V$_{LSR}$, and the distance
are noted at the bottom of each panel.

\noindent
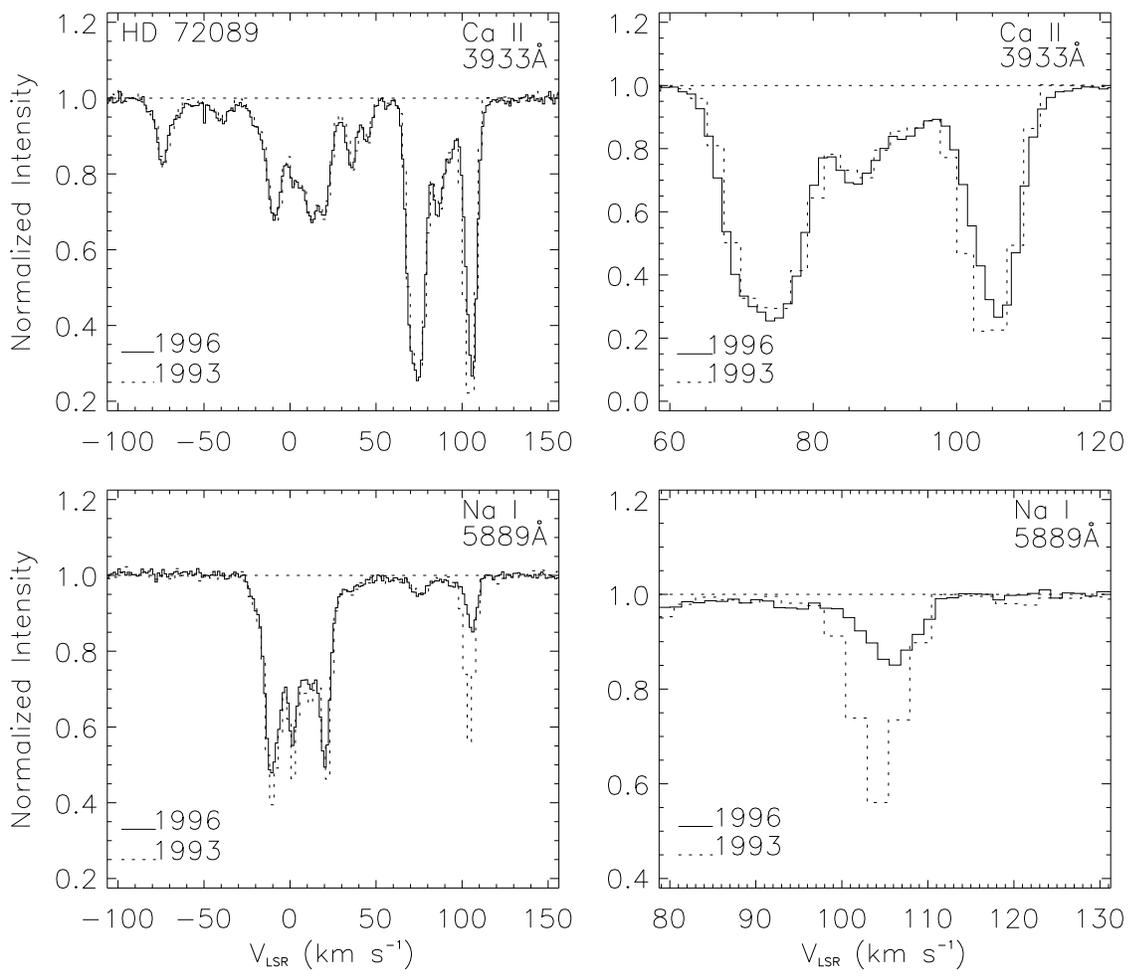
Figure~4.  Comparisons of the \ion{Ca}{2} and \ion{Na}{1} spectra of HD 
72089 observed in February 1993 (dotted lines) and 
January/February 1996 (solid lines).  The left panels show 
the reduced, normalized spectra.  The 
right panels provide close-up views of high velocity 
features that changed between our observations.
Notice the dramatic decrease in the column density 
and displacement of the line center of the \ion{Na}{1} line near +105 
\kms.  

\noindent
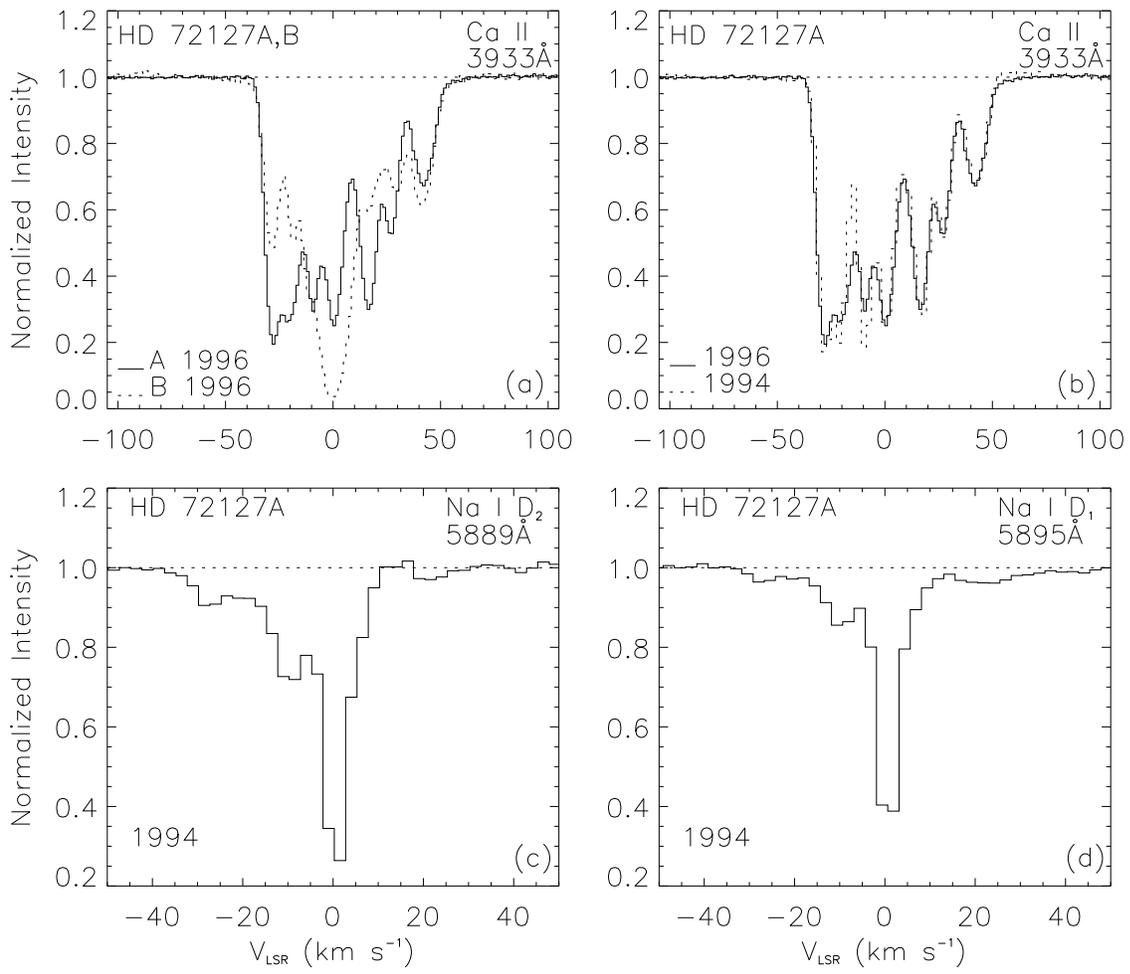
Figure~5.  Panel (a) contains the \ion{Ca}{2} spectra towards the two
components of the visual binary star, HD\,72127A (solid line) and HD\,72127B 
(dotted line).  Panel (b) shows a comparison of the HD\,72127A \ion{Ca}{2}
spectra obtained in 1994 (dotted line) and 1996 (solid line).  Panels (c)
and (d) contain the spectra
of the HD\,72127A \ion{Na}{1} D$_2$ and D$_1$ lines, respectively,
as observed in 1994. 
 
\noindent
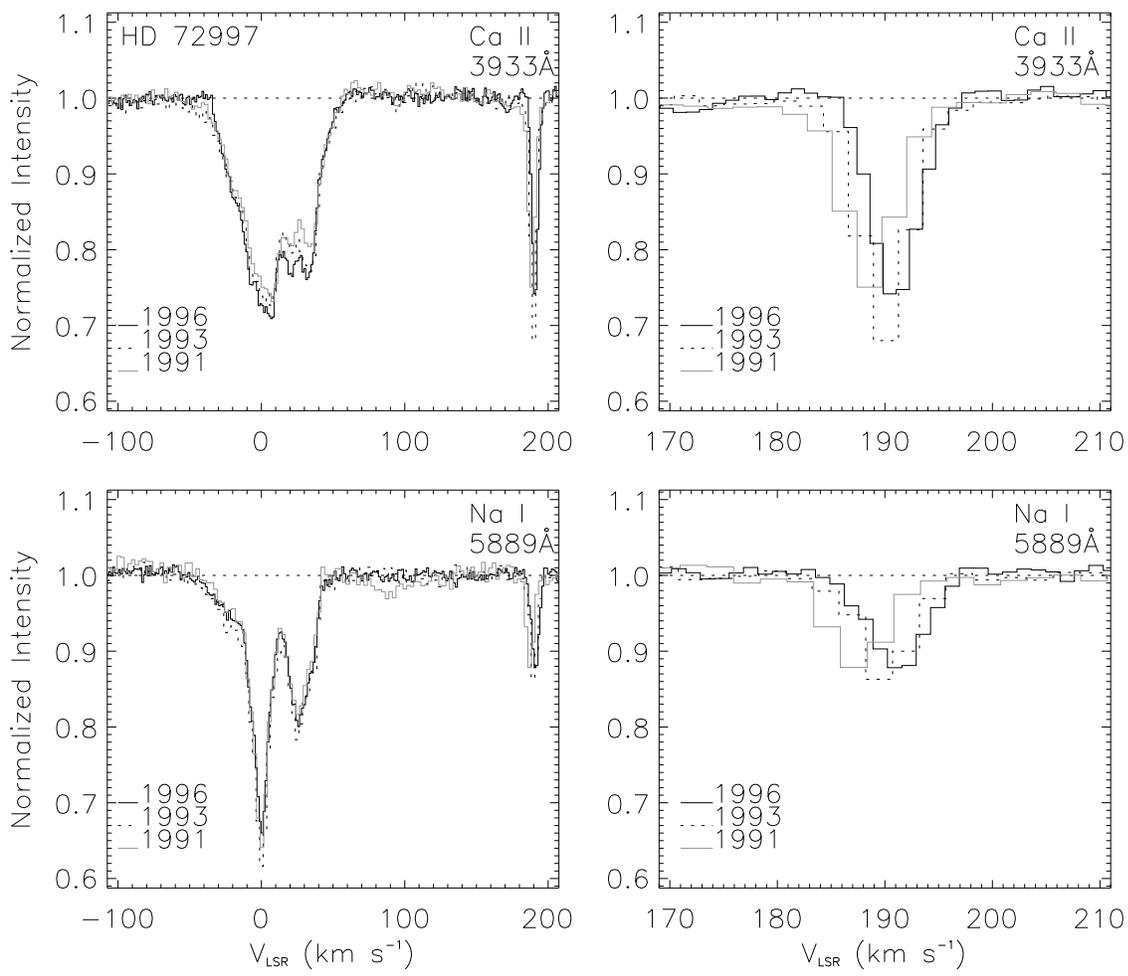
Figure~6.  Comparisons of the \ion{Ca}{2} and \ion{Na}{1} spectra of HD 
72997 observed in December 1991 (gray lines), February 
1993 (dotted lines) and January/February 1996 (black lines).  
The left panels show the reduced, normalized spectra.
The right panels provide close-up views of 
the high velocity feature near +190\kms\ for both \ion{Ca}{2}
and \ion{Na}{1}.  Note the systematic shift of this high velocity 
feature towards higher velocity over the five year period.

\noindent
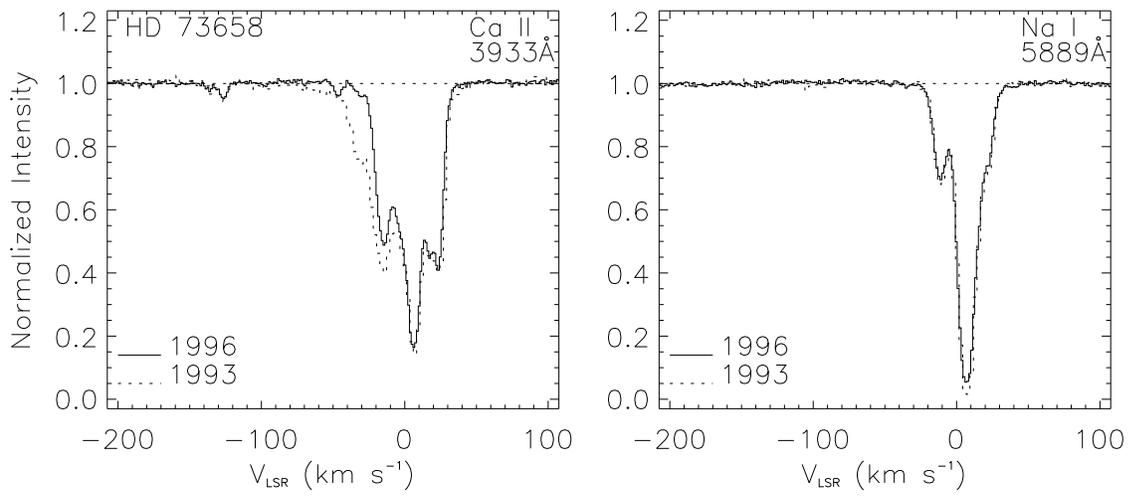
Figure~7.  The left panel shows the \ion{Ca}{2} spectrum of HD\,73658 as
observed in 1993 (dotted line) and 1996 (solid line), while the right panel
depicts the \ion{Na}{1} D$_2$ spectrum as obtained at the same epochs.
The amount of low -- intermediate negative velocity \ion{Ca}{2} gas
decreased between the observations.

\noindent
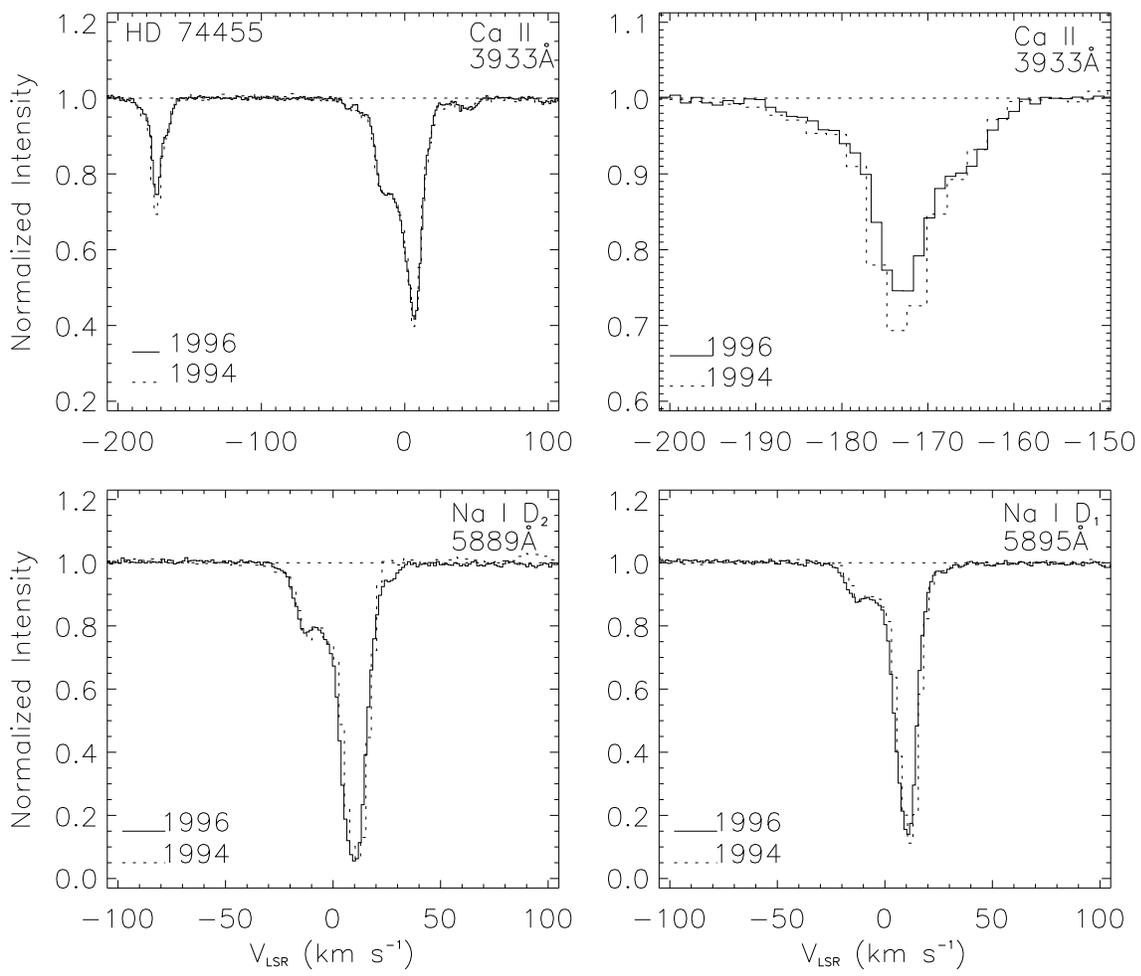
Figure~8.  Spectra of HD\,74455 are shown using dotted lines for the 1994 
observations and solid lines for the 1996 observations.  The upper panels
exhibit \ion{Ca}{2} absorption which varies between the observations in the
high velocity component.  The lower panels illustrate the \ion{Na}{1}
D$_2$ and D$_1$ line profiles which do not vary from 1994 to 1996.

\noindent
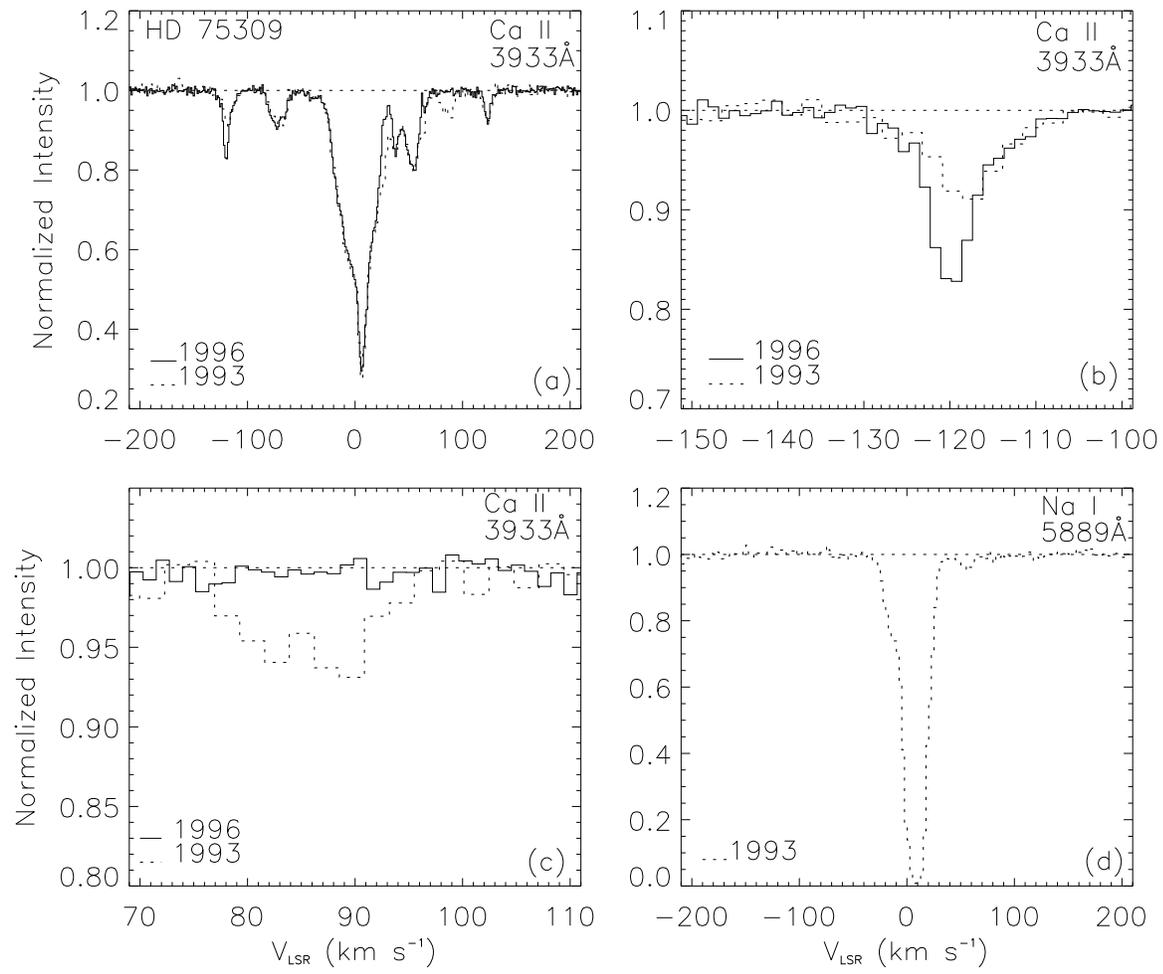
Figure~9.  The \ion{Ca}{2} spectrum of HD\,75309 as observed in
1993 (dotted lines) and 1996 (solid lines) is shown in panel (a).  Panels
(b) and (c) are close up views of two absorption line features that
appear to have changed between observations.  Panel (d) depicts the
\ion{Na}{1} spectrum of HD\,75309 observed in 1993.

\noindent
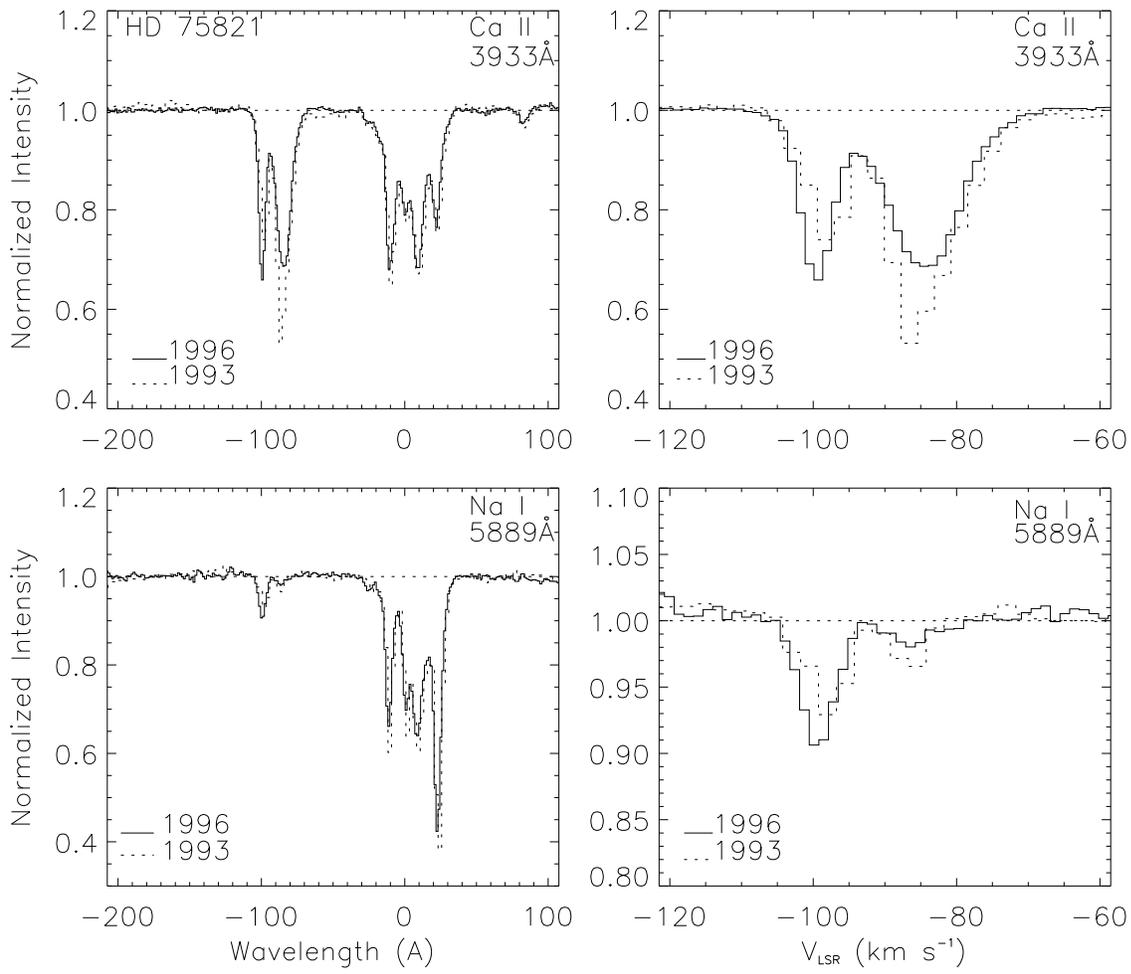
Figure~10.  Comparisons of the \ion{Ca}{2} and \ion{Na}{1} spectra of
HD\,75821 observed in February 1993 (dotted lines) and 
January/February 1996 (solid lines).  The left panels show 
the reduced, normalized spectra of HD\,75821.  The 
right panels provide close-up views of high velocity 
features that occur near $-$85 \kms\ and $-$98 \kms\ for \ion{Ca}{2}
and \ion{Na}{1}, respectively.  These spectra reveal time variability in the 
equivalent widths and velocities of the absorption features.
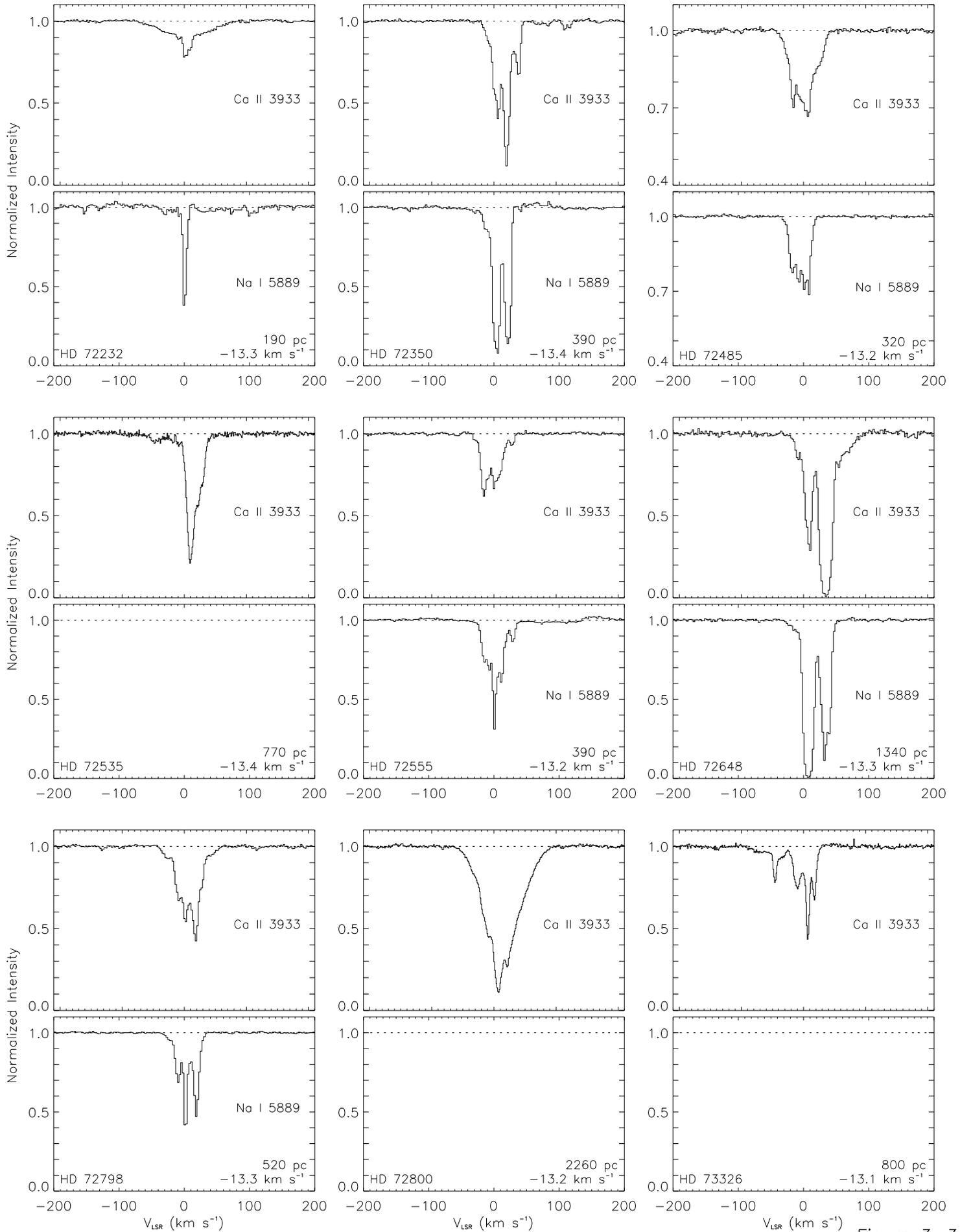
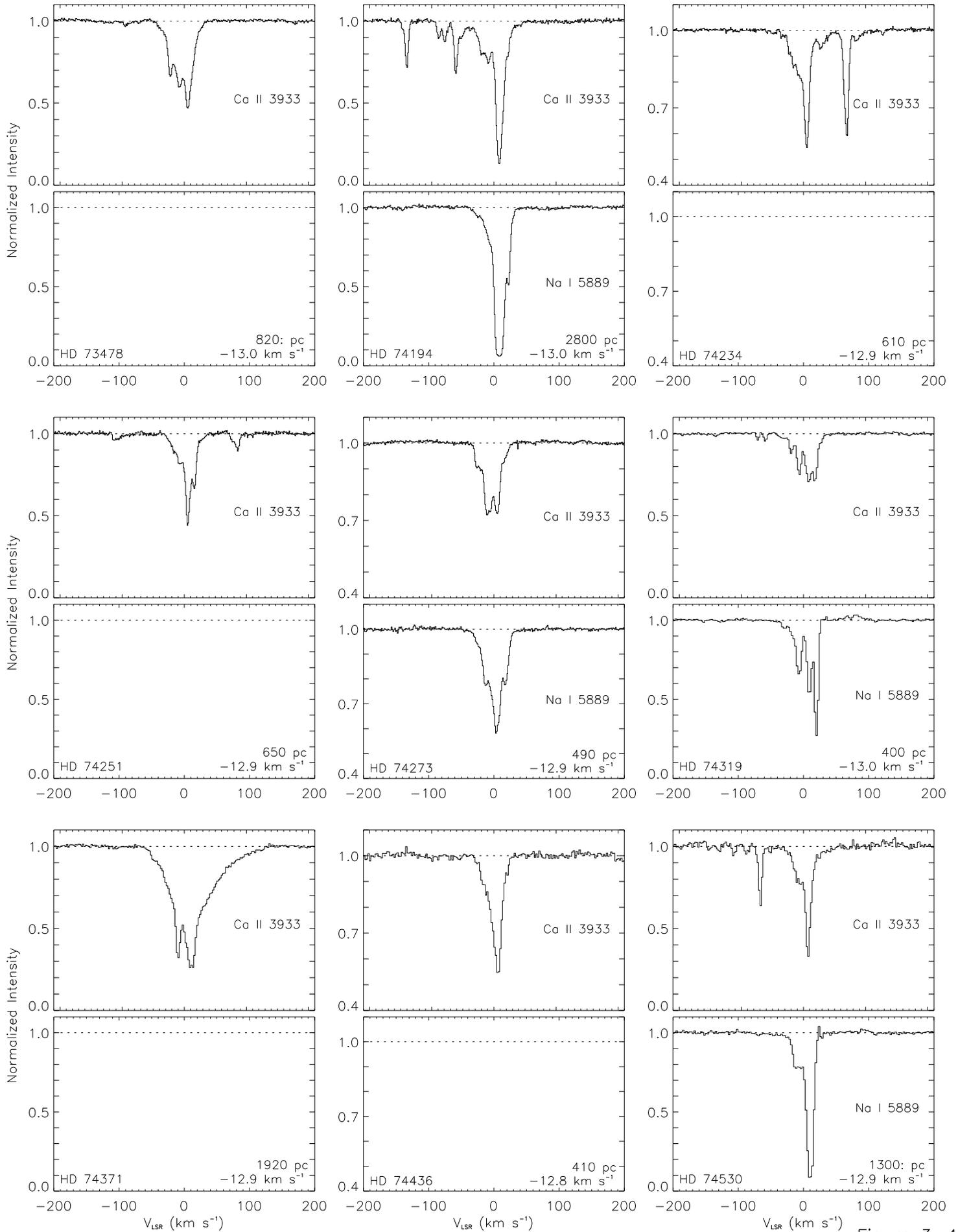
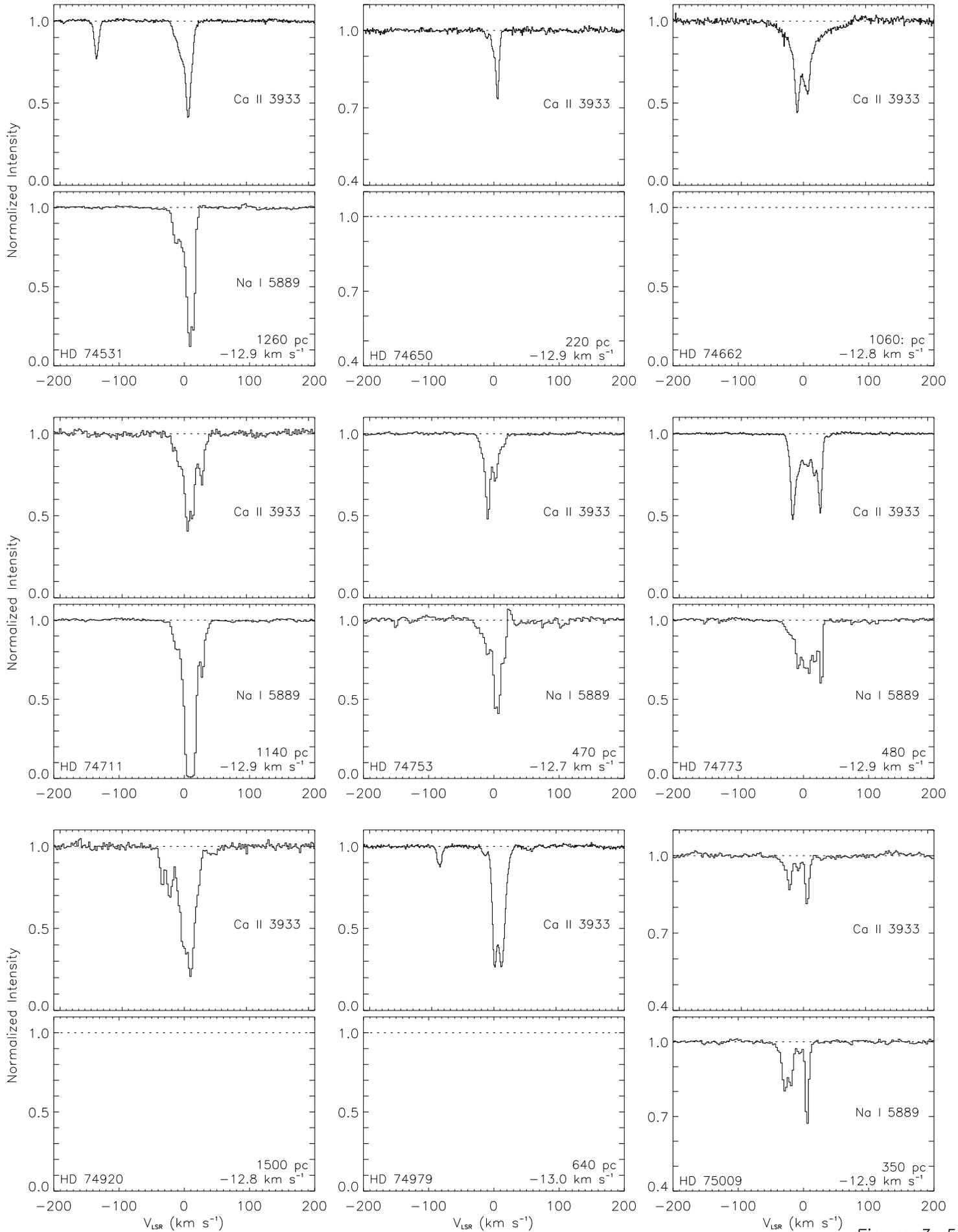
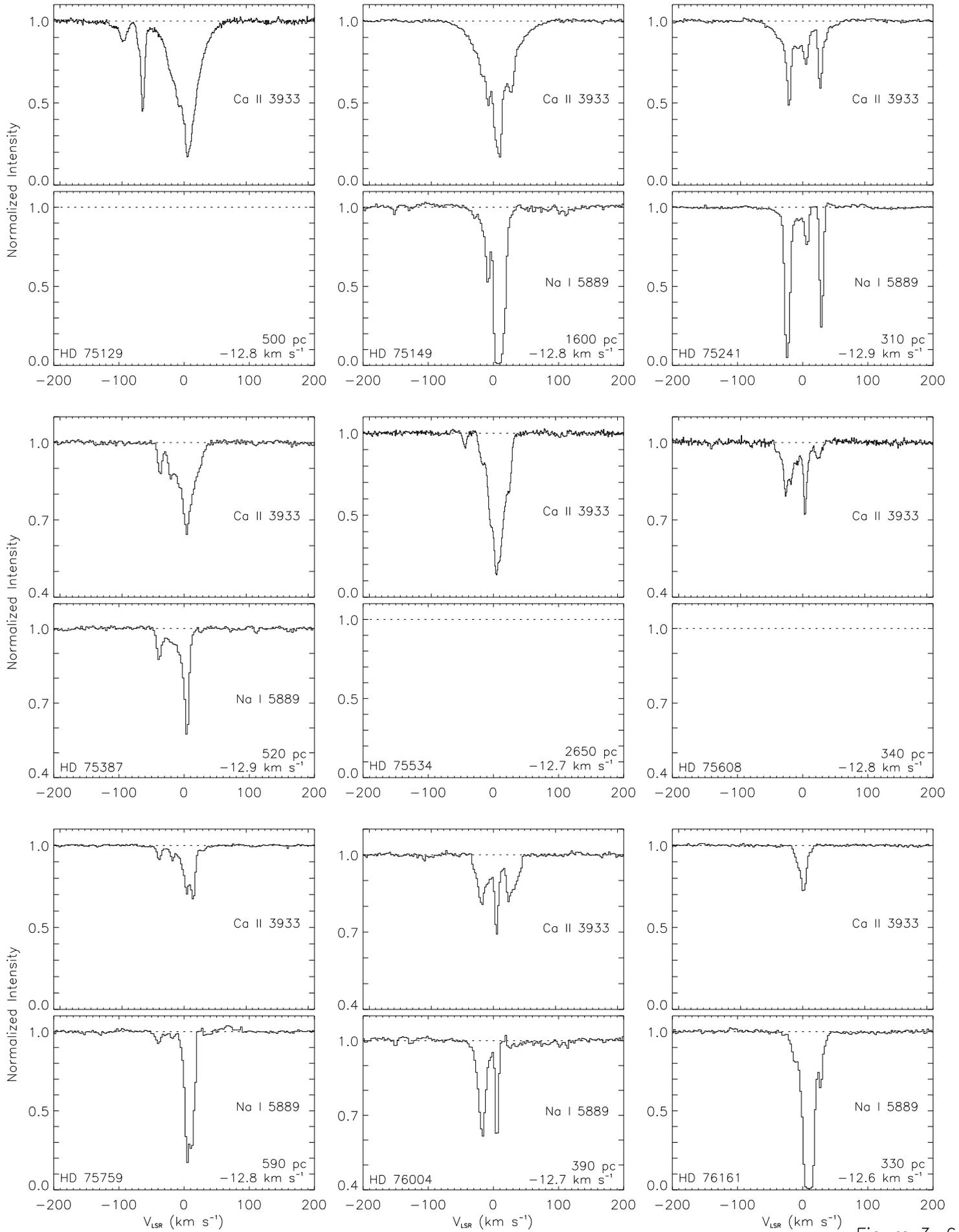
\end{document}


\newcommand{\degr}{$^{\circ}$}
\newcommand{\kms}{\,km\,s$^{-1}$}
\newcommand{\updag}{$^\dagger$}    

\begin{deluxetable}{lcrrccrrcrr}
\tablewidth{0pt}
\setcounter{table}{1}
\tablecaption{\ion{Ca}{2} Absorption Lines} \label{tbl-2}
\tablehead{Star&Obsv. &\multicolumn{1}{c}{V$_{LSR}$}&\multicolumn{1}{c}{W$_\lambda$(K)}&Ref.,&Obsv.&\multicolumn{1}{c}{V$_{LSR}$}&\multicolumn{1}{c}{W$_\lambda$(K)}&Obsv.&\multicolumn{1}{c}{V$_{LSR}$}&\multicolumn{1}{c}{W$_\lambda$(K)}\nl
&Year&\multicolumn{1}{c}{\small (\kms)}&\multicolumn{1}{c}{\small (m\AA)}&Det.&Year&\multicolumn{1}{c}{\small (\kms)}&\multicolumn{1}{c}{\small (m\AA)}&Year&\multicolumn{1}{c}{\small (\kms)}&\multicolumn{1}{c}{\small (m\AA)}}
\startdata
HD 63308& \nodata & \nodata & \nodata & \nodata & 1993 & $-$15 & 29$\pm$1&  \nodata & \nodata & \nodata \nl
& \nodata & \nodata & \nodata & \nodata &1993 & +5& 71$\pm$1&\nodata & \nodata & \nodata \nl
& \nodata & \nodata & \nodata & \nodata &1993 &+16 & 123$\pm$1 &\nodata & \nodata & \nodata \nl
& \nodata & \nodata & \nodata & \nodata &1993 &+28 & 47$\pm$1 &\nodata & \nodata & \nodata \nl
& \nodata & \nodata & \nodata & \nodata &1993 &+45 & 5$\pm$1 &\nodata & \nodata & \nodata \nl
HD 63578&1971-77&$-$14&68&1,p&1993&$-$14&24$\pm$1& \nodata & \nodata & \nodata \nl
&1971-77&+6&86&1,p&1993&+6&34$\pm$1\smallskip& \nodata & \nodata & \nodata  \nl
HD 65814& \nodata & \nodata & \nodata & \nodata &1993&$-$7&37$\pm$2& \nodata & \nodata & \nodata \nl
& \nodata & \nodata & \nodata & \nodata &1993&$-$1&30$\pm$3& \nodata & \nodata & \nodata \nl
& \nodata & \nodata & \nodata & \nodata &1993&+12&160$\pm$2& \nodata & \nodata & \nodata \nl
& \nodata & \nodata & \nodata & \nodata &1993&+23&27$\pm$1& \nodata & \nodata & \nodata \nl
& \nodata & \nodata & \nodata & \nodata &1993&+30&41$\pm$2& \nodata & \nodata & \nodata \smallskip \nl
HD 68217&1971-77&$-$8&43&1,p&1994&$-$8&14$\pm$3& \nodata & \nodata & \nodata \nl
&1971-77&+5&22&1,p &1994&+9&20$\pm$2& \nodata & \nodata & \nodata \smallskip \nl
HD 68243& \nodata & \nodata & \nodata & \nodata &1994&$-$12&8$\pm$1& \nodata & \nodata & \nodata \nl
&1971-77&+1&34&1,p&1994&0&24$\pm$2& \nodata & \nodata & \nodata \nl
&$<$1971&+1&51&4,p& \nodata & \nodata & \nodata & \nodata & \nodata & \nodata\smallskip \nl
HD 68324& \nodata & \nodata & \nodata & \nodata &1994&$-$5&25$\pm$1& \nodata & \nodata & \nodata \nl
& \nodata & \nodata & \nodata & \nodata &1994&+5&18$\pm$1& \nodata & \nodata & \nodata \smallskip \nl
HD 69144& \nodata & \nodata & \nodata & \nodata &1994&$-$8&12$\pm$1& \nodata & \nodata & \nodata \nl
& \nodata & \nodata & \nodata & \nodata &1994&$-$1&8$\pm$3& \nodata & \nodata & \nodata \nl
& \nodata & \nodata & \nodata & \nodata &1994&+5&22$\pm$1& \nodata & \nodata & \nodata \nl
& \nodata & \nodata & \nodata & \nodata &1994&+17&5$\pm$1& \nodata & \nodata & \nodata \smallskip \nl
HD 69302& \nodata & \nodata & \nodata & \nodata & \nodata & \nodata & \nodata &1996&$-$23&5$\pm$1\nl
& \nodata & \nodata & \nodata & \nodata & \nodata & \nodata & \nodata &1996&$-$9&14$\pm$1\nl
& \nodata & \nodata & \nodata & \nodata & \nodata & \nodata & \nodata &1996&$-$1&8$\pm$1\nl
& \nodata & \nodata & \nodata & \nodata & \nodata & \nodata & \nodata &1996&+5&15$\pm$1 \nl
& \nodata & \nodata & \nodata & \nodata & \nodata & \nodata & \nodata &1996&+6&60$\pm$1$^\star$ \nl
& \nodata & \nodata & \nodata & \nodata & \nodata & \nodata & \nodata &1996&+40&6$\pm$1\smallskip \nl

HD 70309& \nodata & \nodata & \nodata & \nodata & \nodata & \nodata & \nodata &1996&$-$28&14$\pm$1\nl
& \nodata & \nodata & \nodata & \nodata & \nodata & \nodata & \nodata &1996&$-$13&21$\pm$1\nl
&1973&0&54&1,p& \nodata & \nodata & \nodata &1996&+1&53$\pm$1\smallskip \nl
HD 70930&1973&$-$15&86&1,p&1993&$-$14&73$\pm$1& \nodata & \nodata & \nodata \nl
&1973&+4&86&1,p& \nodata & \nodata & \nodata & \nodata & \nodata & \nodata\nl
& \nodata & \nodata & \nodata & \nodata &1993&+7&38$\pm$1& \nodata & \nodata & \nodata \nl
& \nodata & \nodata & \nodata & \nodata &1993&+37&12$\pm$1& \nodata & \nodata & \nodata \smallskip \nl
HD 71302& \nodata & \nodata & \nodata & \nodata &1994&$-$20&49$\pm$1& \nodata & \nodata & \nodata \nl
& \nodata & \nodata & \nodata & \nodata &1994&$-$7&19$\pm$1& \nodata & \nodata & \nodata \nl
& \nodata & \nodata & \nodata & \nodata &1994&+3&19$\pm$1& \nodata & \nodata & \nodata \nl
& \nodata & \nodata & \nodata & \nodata &1994&+11&10$\pm$1& \nodata & \nodata & \nodata \nl
& \nodata & \nodata & \nodata & \nodata &1994&+29&11$\pm$1& \nodata & \nodata & \nodata \nl
& \nodata & \nodata & \nodata & \nodata &1994&+44&20$\pm$1& \nodata & \nodata & \nodata \smallskip \nl
HD 71459& \nodata & \nodata & \nodata & \nodata &1993&$-$11&40$\pm$1& \nodata & \nodata & \nodata \nl
& \nodata & \nodata & \nodata & \nodata &1993&+5&53$\pm$1& \nodata & \nodata & \nodata \nl
& \nodata & \nodata & \nodata & \nodata &1993&+30&50$\pm$1& \nodata & \nodata & \nodata \smallskip \nl
\nl
\nl
HD 71609& \nodata & \nodata & \nodata & \nodata & \nodata & \nodata & \nodata &1996&$-$20&35$\pm$1\nl
& \nodata & \nodata & \nodata & \nodata & \nodata & \nodata & \nodata &1996&$-$8&31$\pm$1\nl
& \nodata & \nodata & \nodata & \nodata & \nodata & \nodata & \nodata &1996&+6&107$\pm$1\nl
& \nodata & \nodata & \nodata & \nodata & \nodata & \nodata & \nodata &1996&+17&34$\pm$1\nl
& \nodata & \nodata & \nodata & \nodata & \nodata & \nodata & \nodata &1996&+32&22$\pm$1\nl
& \nodata & \nodata & \nodata & \nodata & \nodata & \nodata & \nodata &1996&+48&15$\pm$1\smallskip \nl
HD 72014& \nodata & \nodata & \nodata & \nodata &1993&$-$19&8$\pm$1&1996&$-$17&11$\pm$1\nl
& \nodata & \nodata & \nodata & \nodata &1993&$-$8&7$\pm$1&1996&$-$7&6$\pm$1\nl
& \nodata & \nodata & \nodata & \nodata &1993&+6&51$\pm$1&1996&+5&41$\pm$1\nl
& \nodata & \nodata & \nodata & \nodata &1993&+15&38$\pm$3&1996&+15&45$\pm$1\nl
& \nodata & \nodata & \nodata & \nodata &1993&+29&50$\pm$1&1996&+29&52$\pm$1\nl
& \nodata & \nodata & \nodata & \nodata &1993&+45&17$\pm$1&1996&+46&16$\pm$1\smallskip \nl
HD 72067&1971-77&0&43&1,p&1994&$-$2&55$\pm$1& \nodata & \nodata & \nodata \nl
&1971-77&+30&54&1,p&1994&+30&32$\pm$1& \nodata & \nodata & \nodata \smallskip \nl
HD 72088& \nodata & \nodata & \nodata & \nodata &1994&$-$56&50$\pm$1& \nodata & \nodata & \nodata \nl
& \nodata & \nodata & \nodata & \nodata &1994&$-$42&25$\pm$1& \nodata & \nodata & \nodata \nl
& \nodata & \nodata & \nodata & \nodata &1994&$-$33&23$\pm$1& \nodata & \nodata & \nodata \nl
& \nodata & \nodata & \nodata & \nodata &1994&$-$21&56$\pm$1& \nodata & \nodata & \nodata \nl
& \nodata & \nodata & \nodata & \nodata &1994&$-$5&206$\pm$1& \nodata & \nodata & \nodata \nl
& \nodata & \nodata & \nodata & \nodata &1994&+8&57$\pm$1& \nodata & \nodata & \nodata \nl
& \nodata & \nodata & \nodata & \nodata &1994&+18&129$\pm$1& \nodata & \nodata & \nodata \nl
& \nodata & \nodata & \nodata & \nodata &1994&+31&126$\pm$1& \nodata & \nodata & \nodata \nl
& \nodata & \nodata & \nodata & \nodata &1994&+108&20$\pm$1& \nodata & \nodata & \nodata \smallskip \nl
HD 72089&$<$1983&$-$73&11&3,p&1993&$-$73&24$\pm$1&1996&$-$74&23$\pm$1\nl
& \nodata & \nodata & \nodata & \nodata &1993&$-$40&9$\pm$1&1996&$-$40&10$\pm$1\nl
&$<$1983&$-$9&21&3,p&1993&$-$8&45$\pm$1&1996&$-$10&53$\pm$1\nl
& \nodata & \nodata & \nodata & \nodata &1993&+4&28$\pm$1&1996&+5&37$\pm$1\nl
&$<$1983&+17&36&3,p&1993&+12&27$\pm$1&1996&+15&39$\pm$1\nl
&1990&+23&197&2,d&1993&+20&40$\pm$1&1996&+22&21$\pm$1\nl
& \nodata & \nodata & \nodata & \nodata &1993&+36&18$\pm$1&1996&+35&21$\pm$1\nl
& \nodata & \nodata & \nodata & \nodata &1993&+45&8$\pm$1&1996&+45&9$\pm$1\nl
&$<$1983&+76&92&3,p&1993&+74&124$\pm$1&1996&+74&126$\pm$1\nl
& \nodata & \nodata & \nodata & \nodata &1993&+87&17$\pm$1&1996&+86&17$\pm$1\nl
& \nodata & \nodata & \nodata & \nodata &1993&+92&22$\pm$1&1996&+93&30$\pm$1\nl
&$<$1983&+108&37&3,p&1993&+105&90$\pm$1&1996&+106&75$\pm$1\nl
&1990&+110&150&2,d& \nodata & \nodata & \nodata & \nodata & \nodata & \nodata \smallskip \nl
HD 72108& \nodata & \nodata & \nodata & \nodata & \nodata & \nodata & \nodata &1996&$-$111&4$\pm$1 \nl
&1973&$-$36&85&1,p& \nodata & \nodata & \nodata & \nodata & \nodata & \nodata \nl
& \nodata & \nodata & \nodata & \nodata & \nodata & \nodata & \nodata &1996&$-$31&13$\pm$1\nl
& \nodata & \nodata & \nodata & \nodata & \nodata & \nodata & \nodata &1996&$-$23&67$\pm$1\nl
&1973&$-$12&35&1,p& \nodata & \nodata & \nodata &1996&$-$12&35$\pm$1\nl
& \nodata & \nodata & \nodata & \nodata & \nodata & \nodata & \nodata &1996&$-$1&20$\pm$1\nl
& \nodata & \nodata & \nodata & \nodata & \nodata & \nodata & \nodata &1996&+5&22$\pm$2\nl
& \nodata & \nodata & \nodata & \nodata & \nodata & \nodata & \nodata &1996&+29&11$\pm$1\smallskip \nl
HD 72127A&1981-88&$-$32&31&6,m& \nodata & \nodata & \nodata & \nodata & \nodata & \nodata \nl
&1981-88&$-$28&63&6,m&1994&$-$28&71$\pm$1&1996&$-$28&92$\pm$1\nl
&1971-77&$-$27&187&1,p& \nodata & \nodata & \nodata & \nodata & \nodata & \nodata \nl
&1981-88&$-$21&76&6,m&1994&$-$21&97$\pm$1&1996&$-$21&126$\pm$1\nl
&1981-88&$-$7&113&6,m&1994&$-$10&76$\pm$1&1996&$-$10&68$\pm$1\nl
&1971-77&$-$6&220&1,p& \nodata & \nodata & \nodata & \nodata & \nodata & \nodata \nl
&1981-88&$-$1&66&6,m&1994&+1&105$\pm$1&1996&+1&111$\pm$1\nl
&1981-88&+4&30&6,m& \nodata & \nodata & \nodata & \nodata & \nodata & \nodata \nl
&1971-77&+13&102&1,p& \nodata & \nodata & \nodata & \nodata & \nodata & \nodata \nl
&1981-88&+14&93&6,m& \nodata & \nodata & \nodata & \nodata & \nodata & \nodata \nl
&1990&+17&582&2,d&1994&+17&104$\pm$1&1996&+16&103$\pm$1\nl

&1971-77&+26&136&1,p&1994&+28&44$\pm$1&1996&+27&47$\pm$1\nl
&1981-88&+27&64&6,m& \nodata & \nodata & \nodata & \nodata & \nodata & \nodata \nl
&1971-77&+41&86&1,p& \nodata & \nodata & \nodata & \nodata & \nodata & \nodata \nl
&1981-88&+42&56&6,m&1994&+42&48$\pm$1&1996&+42&52$\pm$1 \nl
&1981-88&+48&43&6,m& \nodata & \nodata & \nodata & \nodata & \nodata & \nodata \smallskip \nl
HD 72127B& \nodata & \nodata & \nodata & \nodata & \nodata & \nodata & \nodata &1996&$-$28&61$\pm$1\nl
& \nodata & \nodata & \nodata & \nodata & \nodata & \nodata & \nodata &1996&$-$19&45$\pm$1\nl
& \nodata & \nodata & \nodata & \nodata & \nodata & \nodata & \nodata &1996&$-$10&35$\pm$1\nl
& \nodata & \nodata & \nodata & \nodata & \nodata & \nodata & \nodata &1996&0&243$\pm$1\nl
&1990&+1&609&2,d& \nodata & \nodata & \nodata &1996&+17&81$\pm$1\nl
& \nodata & \nodata & \nodata & \nodata & \nodata & \nodata & \nodata &1996&+27&31$\pm$1\nl
& \nodata & \nodata & \nodata & \nodata & \nodata & \nodata & \nodata &1996&+42&64$\pm$1\smallskip \nl
HD 72179& \nodata & \nodata & \nodata & \nodata & \nodata & \nodata & \nodata &1996&$-$33&6$\pm$1\nl
& \nodata & \nodata & \nodata & \nodata & \nodata & \nodata & \nodata &1996&$-$12&66$\pm$1\nl
& \nodata & \nodata & \nodata & \nodata & \nodata & \nodata & \nodata &1996&+1&11$\pm$2\nl
& \nodata & \nodata & \nodata & \nodata & \nodata & \nodata & \nodata &1996&+11&117$\pm$2\nl
& \nodata & \nodata & \nodata & \nodata & \nodata & \nodata & \nodata &1996&+29&10$\pm$1\nl
& \nodata & \nodata & \nodata & \nodata & \nodata & \nodata & \nodata &1996&+35&17$\pm$3\nl
& \nodata & \nodata & \nodata & \nodata & \nodata & \nodata & \nodata &1996&+61&26$\pm$1\nl
& \nodata & \nodata & \nodata & \nodata & \nodata & \nodata & \nodata &1996&+71&65$\pm$3\nl
& \nodata & \nodata & \nodata & \nodata & \nodata & \nodata & \nodata &1996&+78&50$\pm$3\nl
& \nodata & \nodata & \nodata & \nodata & \nodata & \nodata & \nodata &1996&+84&15$\pm$2\nl
& \nodata & \nodata & \nodata & \nodata & \nodata & \nodata & \nodata &1996&+127&19$\pm$1\smallskip \nl
HD 72232&1971-77&\nodata&$<$34&1,p&1994&+2&15$\pm$1& \nodata & \nodata & \nodata\nl
& \nodata & \nodata & \nodata & \nodata &1994&+3&117$\pm$2$^\star$ & \nodata & \nodata & \nodata\smallskip \nl
& \nodata & \nodata & \nodata & \nodata &1994&+10&6$\pm$1& \nodata & \nodata & \nodata\smallskip \nl
HD 72350& \nodata & \nodata & \nodata & \nodata &1994&$-$6&28$\pm$1& \nodata & \nodata & \nodata\nl
& \nodata & \nodata & \nodata & \nodata&1994&0&25$\pm$1& \nodata & \nodata & \nodata\nl
&$<$1983&+6&113&3,p &1994&+7&67$\pm$1& \nodata & \nodata & \nodata\nl
&1971-77&+7&\nodata&1,p& \nodata & \nodata & \nodata & \nodata & \nodata & \nodata \nl
&1971-77&+18&311&1,p& \nodata & \nodata & \nodata& \nodata & \nodata & \nodata\nl
&$<$1983&+21&136&3,p&1994&+20&133$\pm$1& \nodata & \nodata & \nodata\nl
&1990&+26&335&2,d& \nodata & \nodata & \nodata& \nodata & \nodata & \nodata\nl
&$<$1983&+40&37&3,p&1994&+38&48$\pm$1& \nodata & \nodata & \nodata\nl
& \nodata & \nodata & \nodata & \nodata &1994&+114&4$\pm$1& \nodata & \nodata & \nodata\nl
& \nodata & \nodata & \nodata & \nodata &1994&+122&3$\pm$1& \nodata & \nodata & \nodata \smallskip \nl
HD 72485& \nodata & \nodata & \nodata & \nodata &1994&$-$17&11$\pm$2& \nodata & \nodata & \nodata\nl
& \nodata & \nodata & \nodata & \nodata &1994&-2&147$\pm$2$^\star$& \nodata & \nodata & \nodata\nl
& \nodata & \nodata & \nodata & \nodata &1994&+7&16$\pm$1& \nodata & \nodata & \nodata\nl
& \nodata & \nodata & \nodata & \nodata &1994&+27&17$\pm$1& \nodata & \nodata & \nodata \smallskip \nl
HD 72535& \nodata & \nodata & \nodata & \nodata & \nodata & \nodata & \nodata &1996&$-$9&8$\pm$1\nl
& \nodata & \nodata & \nodata & \nodata & \nodata & \nodata & \nodata &1996&+10&142$\pm$1\nl
& \nodata & \nodata & \nodata & \nodata & \nodata & \nodata & \nodata &1996&+21&44$\pm$1\nl
& \nodata & \nodata & \nodata & \nodata & \nodata & \nodata & \nodata &1996&+29&42$\pm$1\smallskip \nl
HD 72555&1973&$-$15&\nodata&1,p&1994&$-$16&56$\pm$1& \nodata & \nodata & \nodata \nl
& \nodata & \nodata & \nodata & \nodata &1994&$-$8&14$\pm$1& \nodata & \nodata & \nodata \nl
& 1973 & +1 & 108 &1,p&1994&0&41$\pm$1& \nodata & \nodata & \nodata \nl
& \nodata & \nodata & \nodata & \nodata &1994&+10&37$\pm$1& \nodata & \nodata & \nodata \nl
& \nodata & \nodata & \nodata & \nodata &1994&+27&11$\pm$1& \nodata & \nodata & \nodata \smallskip \nl
HD 72648& \nodata & \nodata & \nodata & \nodata &1993&$-$9&24$\pm$1& \nodata & \nodata & \nodata \nl
& \nodata & \nodata & \nodata & \nodata &1993&+10&103$\pm$2& \nodata & \nodata & \nodata \nl
&1990&+35&623&2,d&1993&+36&321$\pm$1& \nodata & \nodata & \nodata \nl
& \nodata & \nodata & \nodata & \nodata &1993&+57&11$\pm$3& \nodata & \nodata & \nodata \smallskip \nl
HD 72798 & \nodata & \nodata & \nodata & \nodata &1994&$-$25&19$\pm$1& \nodata & \nodata & \nodata \nl
& \nodata & \nodata & \nodata & \nodata &1994&$-$10&40$\pm$1& \nodata & \nodata & \nodata \nl
& \nodata & \nodata & \nodata & \nodata &1994&+2&82$\pm$3& \nodata & \nodata & \nodata \nl
&1971-77&+15&274&1,p &1994&+18&96$\pm$2 & \nodata & \nodata & \nodata  \nl
& \nodata & \nodata & \nodata & \nodata &1994&+29&13$\pm$2 & \nodata & \nodata & \nodata \nl
& \nodata & \nodata & \nodata & \nodata &1994&+39&20$\pm$2 & \nodata & \nodata & \nodata \smallskip \nl
HD 72800& \nodata & \nodata & \nodata & \nodata &\nodata & \nodata & \nodata &1996 &$-$35 & 26$\pm$1 \nl
& \nodata & \nodata & \nodata & \nodata &\nodata & \nodata & \nodata &1996 &$-$10 & 106$\pm$1 \nl
& \nodata & \nodata & \nodata & \nodata &\nodata & \nodata & \nodata &1996 &+7 & 124$\pm$1 \nl
& \nodata & \nodata & \nodata & \nodata &\nodata & \nodata & \nodata &1996 &+21 & 90$\pm$2 \nl
& \nodata & \nodata & \nodata & \nodata &\nodata & \nodata & \nodata &1996 &+21 & 470$\pm$2$^\star$ \smallskip \nl
HD 72997& \nodata & \nodata & \nodata & \nodata &1993&$-$20&37$\pm$1 &1996&$-$18&32$\pm$1 \nl
& \nodata & \nodata & \nodata & \nodata &1993&$-$3&52$\pm$1&1996&$-$5&42$\pm$1\nl
& \nodata & \nodata & \nodata & \nodata &1993&+9&35$\pm$1&1996&+7&46$\pm$1\nl
& \nodata & \nodata & \nodata & \nodata &1993&+20&12$\pm$1&1996&+19&20$\pm$1\nl
& \nodata & \nodata & \nodata & \nodata &1993&+32&63$\pm$1&1996&+31&63$\pm$1\nl
& \nodata & \nodata & \nodata & \nodata &1993&+190&23$\pm$1&1996&+191&19$\pm$1\smallskip \nl
HD 73326& \nodata & \nodata & \nodata & \nodata & \nodata & \nodata & \nodata &1996&$-$46&14$\pm$1\nl
& \nodata & \nodata & \nodata & \nodata & \nodata & \nodata & \nodata &1996&$-$37&18$\pm$1\nl
& \nodata & \nodata & \nodata & \nodata & \nodata & \nodata & \nodata &1996&$-$10&50$\pm$1\nl
& \nodata & \nodata & \nodata & \nodata & \nodata & \nodata & \nodata &1996&0&7$\pm$2\nl
& \nodata & \nodata & \nodata & \nodata & \nodata & \nodata & \nodata &1996&+7&48$\pm$1\nl
& \nodata & \nodata & \nodata & \nodata & \nodata & \nodata & \nodata &1996&+17&36$\pm$1\smallskip \nl
HD 73478& \nodata & \nodata & \nodata & \nodata & \nodata & \nodata & \nodata &1996&$-$36&10$\pm$1\nl
& \nodata & \nodata & \nodata & \nodata & \nodata & \nodata & \nodata &1996&$-$22&44$\pm$1\nl
& \nodata & \nodata & \nodata & \nodata & \nodata & \nodata & \nodata &1996&$-$8&73$\pm$1\nl
& \nodata & \nodata & \nodata & \nodata & \nodata & \nodata & \nodata &1996&+5&70$\pm$1\nl
& \nodata & \nodata & \nodata & \nodata & \nodata & \nodata & \nodata &1996&+13&51$\pm$2\smallskip \nl
HD 73658& \nodata & \nodata & \nodata & \nodata &1993&-126&5$\pm$1&1996&-126&4$\pm$1\nl
& \nodata & \nodata & \nodata & \nodata &1993&$-$33&41$\pm$1&1996&$-$32&4$\pm$2 \nl
& \nodata & \nodata & \nodata & \nodata &1993&$-$15&144$\pm$1&1996&$-$15&91$\pm$1\nl
& \nodata & \nodata & \nodata & \nodata &1993&$-$2&51$\pm$1&1996&-3&42$\pm$1 \nl
& \nodata & \nodata & \nodata & \nodata &1993&+7&126$\pm$1&1996&+6&140$\pm$1\nl
& \nodata & \nodata & \nodata & \nodata&1993&+18&68$\pm$1&1996&+17&36$\pm$2\nl
&1990&+21&507&2,d& \nodata & \nodata & \nodata & \nodata& \nodata & \nodata \nl
& \nodata & \nodata & \nodata & \nodata &1993&+25&58$\pm$1&1996&+24&67$\pm$1\smallskip \nl
HD 74194& \nodata & \nodata & \nodata & \nodata &1993&$-$141&30$\pm$1&1996&$-$140&23$\pm$1\nl
& \nodata & \nodata & \nodata & \nodata &1993&$-$88&11$\pm$1&1996&$-$89&9$\pm$1\nl
& \nodata & \nodata & \nodata & \nodata &1993&$-$78&14$\pm$1&1996&$-$79&12$\pm$1\nl
&1973&$-$63&172&1,p& 1993&$-$61&31$\pm$1&1996&$-$61&20$\pm$1\nl
& \nodata & \nodata & \nodata & \nodata &1993&$-$49&10$\pm$1&1996&$-$55&31$\pm$1\nl
& \nodata & \nodata & \nodata & \nodata &1993&$-$17&50$\pm$1&1996&$-$20&50$\pm$1\nl
& \nodata & \nodata & \nodata & \nodata &1993&$-$7&21$\pm$1&1996&$-$8&26$\pm$1\nl
&1973&+8&252&1,p&1993&+9&133$\pm$1&1996&+8&134$\pm$1\smallskip \nl
HD 74234& \nodata & \nodata & \nodata & \nodata & \nodata & \nodata & \nodata &1996&$-$17&27$\pm$1\nl
&$<$1983&$-$3&20&3,p& \nodata & \nodata & \nodata &1996&$-$6&18$\pm$1\nl
&1973&+2&216&1,p& \nodata & \nodata & \nodata & \nodata & \nodata & \nodata \nl
&$<$1983&+6&180&3,p& \nodata & \nodata & \nodata &1996&+5&64$\pm$1\nl
& \nodata & \nodata & \nodata & \nodata & \nodata & \nodata & \nodata &1996&+28&15$\pm$1\nl
& \nodata & \nodata & \nodata & \nodata & \nodata & \nodata & \nodata &1996&+69&47$\pm$1\nl
& \nodata & \nodata & \nodata & \nodata & \nodata & \nodata & \nodata &1996&+85&6$\pm$1\smallskip \nl
HD 74251& \nodata & \nodata & \nodata & \nodata & \nodata & \nodata & \nodata &1996&$-$108&8$\pm$1\nl
& \nodata & \nodata & \nodata & \nodata & \nodata & \nodata & \nodata &1996&$-$5&41$\pm$1\nl
& \nodata & \nodata & \nodata & \nodata & \nodata & \nodata & \nodata &1996&+6&56$\pm$1\nl
&1973&+8&216&1,p& \nodata & \nodata & \nodata & \nodata & \nodata & \nodata \nl
& \nodata & \nodata & \nodata & \nodata & \nodata & \nodata & \nodata &1996&+16&40$\pm$1\nl
& \nodata & \nodata & \nodata & \nodata & \nodata & \nodata & \nodata &1996&+78&5$\pm$1\nl
& \nodata & \nodata & \nodata & \nodata & \nodata & \nodata & \nodata &1996&+86&9$\pm$1\smallskip \nl
HD 74273& \nodata & \nodata & \nodata & \nodata &1994&$-$25&21$\pm$1&1996&$-$27&11$\pm$1\nl
& \nodata & \nodata & \nodata & \nodata &1994&$-$9&51$\pm$1&1996&$-$9&66$\pm$1\nl
&$<$1971&$-$1&117&4,p& \nodata & \nodata & \nodata & \nodata & \nodata & \nodata \nl
&1971&+2&172&1,p& \nodata & \nodata & \nodata & \nodata & \nodata & \nodata \nl
& \nodata & \nodata & \nodata & \nodata &1994&+5&43$\pm$1&1996&+6&34$\pm$1\smallskip \nl
HD 74319& \nodata & \nodata & \nodata & \nodata &1994&$-$19&15$\pm$1& \nodata & \nodata & \nodata \nl
& \nodata & \nodata & \nodata & \nodata &1994&$-$7&29$\pm$1& \nodata & \nodata & \nodata \nl
& \nodata & \nodata & \nodata & \nodata &1994&+8&41$\pm$1& \nodata & \nodata & \nodata \nl
& \nodata & \nodata & \nodata & \nodata &1994&+19&38$\pm$1& \nodata & \nodata & \nodata \smallskip \nl
HD 74371& \nodata & \nodata & \nodata & \nodata&1994&$-$10&42$\pm$2& \nodata & \nodata & \nodata \nl
&$<$1971&$-$5&304&4,p& \nodata & \nodata & \nodata & \nodata& \nodata & \nodata \nl
&1971&$-$2&\nodata&1,p &1994 & +2 & 286$\pm$3$^\bullet$ & \nodata & \nodata & \nodata \nl
&$<$1971&+10&330&4,p&1994&+11&96$\pm$2& \nodata & \nodata & \nodata \nl
&1971&+13&\nodata&1,p\nodata & \nodata & \nodata & \nodata& \nodata & \nodata \nl
& \nodata & \nodata & \nodata & \nodata& 1994 & +29 & 301$\pm$3$^\bullet$ & \nodata & \nodata & \nodata \smallskip \nl
HD 74436& \nodata & \nodata & \nodata & \nodata &1994&$-$7&57$\pm$1& \nodata & \nodata & \nodata \nl
& \nodata & \nodata & \nodata & \nodata &1994&+7&80$\pm$1& \nodata & \nodata & \nodata \smallskip \nl
HD 74455& \nodata & \nodata & \nodata &\nodata&1994&$-$183&6$\pm$1&1996&$-$182&5$\pm$1 \nl
&$<$1971&$-$174&36&4,p&1994&$-$173&39$\pm$1&1996&$-$173&21$\pm$1\nl
&1971-73&$-$173&43&1,p& \nodata & \nodata & \nodata & \nodata & \nodata & \nodata \nl
&$<$1971&$-$166&\nodata&5,p& 1994 & $-$166&6$\pm$1 &1996&$-$166&8$\pm$1\nl
&1990&$-$159&28&2,d& \nodata & \nodata & \nodata & \nodata & \nodata & \nodata \nl
&$<$1971&$-$70&\nodata&5,p& \nodata & \nodata & \nodata & \nodata & \nodata & \nodata \nl
&$<$1971&$-$24&\nodata&5,p& \nodata & \nodata & \nodata & \nodata & \nodata & \nodata \nl
&1971-73&$-$6&68&1,p&1994&$-$13&56$\pm$1&1996&$-$14&55$\pm$1\nl
& \nodata & \nodata & \nodata & \nodata &1994&$-$4&14$\pm$1&1996&$-$2&34$\pm$1\nl
&$<$1971&+5&204&4,p&1994&+6&99$\pm$1&1996&+7&88$\pm$1\nl
&$<$1971&+5&\nodata&5,p& \nodata & \nodata & \nodata & \nodata & \nodata & \nodata \nl
&1971-73&+7&108&1,p&\nodata & \nodata & \nodata & \nodata & \nodata & \nodata \nl
&1990&+18&216&2,d& \nodata & \nodata & \nodata & \nodata & \nodata & \nodata \smallskip \nl
HD 74530& \nodata & \nodata & \nodata & \nodata &1994&$-$69&30$\pm$1& \nodata & \nodata & \nodata \nl
& \nodata & \nodata & \nodata & \nodata &1994&$-$5&67$\pm$1& \nodata & \nodata & \nodata \nl
& \nodata & \nodata & \nodata & \nodata &1994&+7&61$\pm$1& \nodata & \nodata & \nodata \nl
& \nodata & \nodata & \nodata & \nodata &1994&+13&41$\pm$1& \nodata & \nodata & \nodata \smallskip \nl
HD 74531&1973&$-$152&34&1,p & \nodata & \nodata & \nodata & \nodata & \nodata & \nodata \nl
& \nodata & \nodata & \nodata & \nodata &1994&$-$141&34$\pm$1&1996&$-$141&25$\pm$1\nl
&1973&+5&216&1,p &1994&+7&70$\pm$1&1996&+7&94$\pm$1 \smallskip \nl
HD 74650& \nodata & \nodata & \nodata & \nodata & \nodata & \nodata & \nodata &1996&$-$1&8$\pm$1\nl
& \nodata & \nodata & \nodata & \nodata & \nodata & \nodata & \nodata &1996&+6&21$\pm$1\smallskip \nl
HD 74662& \nodata & \nodata & \nodata & \nodata & \nodata & \nodata & \nodata &1996&$-$19&6$\pm$2\nl
& \nodata & \nodata & \nodata & \nodata & \nodata & \nodata & \nodata &1996&$-$10&54$\pm$2\nl
& \nodata & \nodata & \nodata & \nodata & \nodata & \nodata & \nodata &1996&$-$2&14$\pm$4\nl
& \nodata & \nodata & \nodata & \nodata & \nodata & \nodata & \nodata &1996&+2&152$\pm$3$^\bullet$ \nl
& \nodata & \nodata & \nodata & \nodata & \nodata & \nodata & \nodata &1996&+6&47$\pm$3\smallskip \nl
HD 74711& \nodata & \nodata & \nodata & \nodata &1993&$-$4&55$\pm$2& \nodata & \nodata & \nodata \nl
& \nodata & \nodata & \nodata & \nodata &1993&+5&59$\pm$2& \nodata & \nodata & \nodata \nl
& \nodata & \nodata & \nodata & \nodata &1993&+14&55$\pm$2& \nodata & \nodata & \nodata \nl
& \nodata & \nodata & \nodata & \nodata &1993&+27&46$\pm$1& \nodata & \nodata & \nodata \smallskip \nl
HD 74753&1971&$-$8&172&1,p&1994&$-$9&53$\pm$1& \nodata & \nodata & \nodata \nl
&$<$1971&$-$8&122&4,p& \nodata & \nodata & \nodata & \nodata & \nodata & \nodata \nl
& \nodata & \nodata & \nodata & \nodata &1994&+2&39$\pm$1& \nodata & \nodata & \nodata \smallskip \nl
HD 74773& \nodata & \nodata & \nodata & \nodata &1994&$-$18&48$\pm$1&1996&$-$17&59$\pm$1\nl
& \nodata & \nodata & \nodata & \nodata &1994&$-$11&43$\pm$1&1996&$-$9&26$\pm$1\nl
& \nodata & \nodata & \nodata & \nodata &1994&+5&49$\pm$2&1996&+5&50$\pm$1\nl
& \nodata & \nodata & \nodata & \nodata &1994&+18&26$\pm$2&1996&+18&25$\pm$1\nl
& \nodata & \nodata & \nodata & \nodata &1994&+27&46$\pm$1&1996&+27&44$\pm$1\smallskip \nl
HD 74920& \nodata & \nodata & \nodata & \nodata &1993&$-$36&23$\pm$2& \nodata & \nodata & \nodata \nl
& \nodata & \nodata & \nodata & \nodata &1993&$-$23&50$\pm$2& \nodata & \nodata & \nodata \nl
&$<$1971&$-$8&122&4,p&1993&$-$9&34$\pm$3& \nodata & \nodata & \nodata \nl
& \nodata & \nodata & \nodata & \nodata &1993&$-$3&42$\pm$1& \nodata & \nodata & \nodata \nl
&$<$1983&+6&40&3,p&1993&+2&35$\pm$4& \nodata & \nodata & \nodata \nl
& \nodata & \nodata & \nodata & \nodata &1993&+10&134$\pm$2& \nodata & \nodata & \nodata \nl
& \nodata & \nodata & \nodata & \nodata &1993&+21&40$\pm$3& \nodata & \nodata & \nodata \nl
&\nodata & \nodata &\nodata& \nodata &1993&+45&14$\pm$1& \nodata & \nodata & \nodata \smallskip \nl
HD 74979& \nodata & \nodata & \nodata & \nodata & \nodata & \nodata & \nodata &1996&$-$87&15$\pm$1\nl
& \nodata & \nodata & \nodata & \nodata & \nodata & \nodata & \nodata &1996&$-$13&9$\pm$1\nl
& \nodata & \nodata & \nodata & \nodata & \nodata & \nodata & \nodata &1996&+1&73$\pm$1\nl
& \nodata & \nodata & \nodata & \nodata & \nodata & \nodata & \nodata &1996&+12&146$\pm$1\smallskip \nl
HD 75009& \nodata & \nodata & \nodata & \nodata &1994&$-$22&10$\pm$1& \nodata & \nodata & \nodata \nl
& \nodata & \nodata & \nodata & \nodata &1994&$-$8&7$\pm$1& \nodata & \nodata & \nodata \nl
& \nodata & \nodata & \nodata & \nodata &1994&+6&19$\pm$1& \nodata & \nodata & \nodata \smallskip \nl
HD 75129& \nodata & \nodata & \nodata & \nodata & \nodata & \nodata & \nodata &1996&$-$98&26$\pm$1\nl
& \nodata & \nodata & \nodata & \nodata & \nodata & \nodata & \nodata &1996&$-$67&62$\pm$1\nl
& \nodata & \nodata & \nodata & \nodata & \nodata & \nodata & \nodata &1996&+2&396$\pm$2$^\bullet$\nl
& \nodata & \nodata & \nodata & \nodata & \nodata & \nodata & \nodata &1996&+7&109$\pm$1\smallskip \nl
HD 75149& \nodata & \nodata & \nodata & \nodata &1994&$-$24&59$\pm$2$^\bullet$& \nodata & \nodata & \nodata \nl
& \nodata & \nodata & \nodata & \nodata &1994&$-$19&14$\pm$1& \nodata & \nodata & \nodata \nl
& \nodata & \nodata & \nodata & \nodata &1994&$-$9&43$\pm$1& \nodata & \nodata & \nodata \nl
& \nodata & \nodata & \nodata & \nodata &1994&+8&163$\pm$1& \nodata & \nodata & \nodata \nl
&$<$1971&+8&450&4,p& \nodata & \nodata & \nodata & \nodata & \nodata & \nodata \nl
& \nodata & \nodata & \nodata & \nodata &1994&+13&213$\pm$1$^\star$& \nodata & \nodata & \nodata \nl
& \nodata & \nodata & \nodata & \nodata &1994&+28&52$\pm$1& \nodata & \nodata & \nodata \smallskip \nl
HD 75241& \nodata & \nodata & \nodata & \nodata &1994&$-$52&11$\pm$1& \nodata & \nodata & \nodata \nl
& \nodata & \nodata & \nodata & \nodata &1994&$-$40&12$\pm$1& \nodata & \nodata & \nodata \nl
& \nodata & \nodata & \nodata & \nodata &1994&$-$27&31$\pm$1& \nodata & \nodata & \nodata \nl
& \nodata & \nodata & \nodata & \nodata &1994&$-$22&41$\pm$1& \nodata & \nodata & \nodata \nl
& \nodata & \nodata & \nodata & \nodata &1994&$-$9&45$\pm$1& \nodata & \nodata & \nodata \nl
& \nodata & \nodata & \nodata & \nodata &1994&+6&30$\pm$1& \nodata & \nodata & \nodata \nl
& \nodata & \nodata & \nodata & \nodata &1994&+17&6$\pm$1& \nodata & \nodata & \nodata \nl
& \nodata & \nodata & \nodata & \nodata &1994&+28&44$\pm$1& \nodata & \nodata & \nodata \nl
& \nodata & \nodata & \nodata & \nodata &1994&+37&8$\pm$1& \nodata & \nodata & \nodata \nl
& \nodata & \nodata & \nodata & \nodata &1994&+47&9$\pm$1& \nodata & \nodata & \nodata \smallskip \nl
HD 75309& \nodata & \nodata & \nodata & \nodata &1993&$-$118&12$\pm$1&1996&$-$120&16$\pm$1\nl
& \nodata & \nodata & \nodata & \nodata &1993&$-$74&12$\pm$2&1996&$-$72&20$\pm$1\nl
& \nodata & \nodata & \nodata & \nodata &1993&$-$6&130$\pm$1&1996&$-$5&158$\pm$1\nl
& \nodata & \nodata & \nodata & \nodata &1993&+7&53$\pm$2&1996&+7&71$\pm$1\nl
& \nodata & \nodata & \nodata & \nodata &1993&+14&146$\pm$2&1996&+17&60$\pm$1\nl
& \nodata & \nodata & \nodata & \nodata &1993&+38&7$\pm$2&1996&+38&16$\pm$1\nl
& \nodata & \nodata & \nodata & \nodata &1993&+54&47$\pm$1&1996&+53&39$\pm$1\nl
& \nodata & \nodata & \nodata & \nodata &1993&+81&4$\pm$2&1996&+81&$<$1 \nl
& \nodata & \nodata & \nodata & \nodata &1993&+89&7$\pm$2&1996&+89&$<$1 \nl
& \nodata & \nodata & \nodata & \nodata &1993&+123&9$\pm$1&1996&+123&7$\pm$1\smallskip \nl
HD 75387& \nodata & \nodata & \nodata & \nodata &1994&$-$39&14$\pm$1& \nodata & \nodata & \nodata \nl
& \nodata & \nodata & \nodata & \nodata &1994&$-$20&25$\pm$1& \nodata & \nodata & \nodata \nl
& \nodata & \nodata & \nodata & \nodata &1994&$-$9&14$\pm$1& \nodata & \nodata & \nodata \nl
& \nodata & \nodata & \nodata & \nodata &1994&+3&60$\pm$1& \nodata & \nodata & \nodata \nl
& \nodata & \nodata & \nodata & \nodata &1994&+17&44$\pm$1$^\star$& \nodata & \nodata & \nodata \smallskip \nl
HD 75534& \nodata & \nodata & \nodata & \nodata & \nodata & \nodata & \nodata &1996&$-$46&8$\pm$1\nl
& \nodata & \nodata & \nodata & \nodata & \nodata & \nodata & \nodata &1996&$-$17&32$\pm$1\nl
& \nodata & \nodata & \nodata & \nodata & \nodata & \nodata & \nodata &1996&$-$5&62$\pm$1\nl
& \nodata & \nodata & \nodata & \nodata & \nodata & \nodata & \nodata &1996&+5&128$\pm$1\nl
& \nodata & \nodata & \nodata & \nodata & \nodata & \nodata & \nodata &1996&+13&72$\pm$1\nl
& \nodata & \nodata & \nodata & \nodata & \nodata & \nodata & \nodata &1996&+23&62$\pm$1\smallskip \nl
HD 75608& \nodata & \nodata & \nodata & \nodata & \nodata & \nodata & \nodata &1996&$-$37&8$\pm$1\nl
& \nodata & \nodata & \nodata & \nodata & \nodata & \nodata & \nodata &1996&$-$27&21$\pm$2\nl
& \nodata & \nodata & \nodata & \nodata & \nodata & \nodata & \nodata &1996&$-$19&14$\pm$2\nl
& \nodata & \nodata & \nodata & \nodata & \nodata & \nodata & \nodata &1996&$-$9&12$\pm$1\nl
& \nodata & \nodata & \nodata & \nodata & \nodata & \nodata & \nodata &1996&+4&26$\pm$1\nl
& \nodata & \nodata & \nodata & \nodata & \nodata & \nodata & \nodata &1996&+25&12$\pm$1\smallskip \nl
HD 75759& \nodata & \nodata & \nodata & \nodata &1994&$-$40&11$\pm$1& \nodata & \nodata & \nodata \nl
& \nodata & \nodata & \nodata & \nodata &1994&$-$19&12$\pm$1& \nodata & \nodata & \nodata \nl
&1971-77&+7&108&1,p&1994&+5&45$\pm$1& \nodata & \nodata & \nodata \nl
& \nodata & \nodata & \nodata & \nodata &1994&+15&32$\pm$1& \nodata & \nodata & \nodata \smallskip \nl
HD 75821&$<$1971&$-$126&39&5,p& \nodata & \nodata & \nodata & \nodata & \nodata & \nodata \nl
& \nodata & \nodata & \nodata & \nodata &1993&$-$98&21$\pm$1&1996&$-$99&26$\pm$1\nl
&$<$1971&$-$88&73&4,p& \nodata & \nodata & \nodata & \nodata & \nodata & \nodata \nl
&$<$1971&$-$89&59&5,p& \nodata & \nodata & \nodata & \nodata & \nodata & \nodata \nl
&1971&$-$86&108&1,p&1993&$-$85&66$\pm$1&1996&$-$85&53$\pm$1\nl
&1971&$-$9&54&1,p&1993&$-$9&30$\pm$1&1996&$-$10&30$\pm$1\nl
&$<$1971&+3&140&5,p &1993&+1&23$\pm$1&1996&0&22$\pm$1\nl
&$<$1971&+4&95&4,p& \nodata & \nodata & \nodata & \nodata & \nodata & \nodata \nl
&1971&+9&108&1,p&1993&+10&41$\pm$1&1996&+9&39$\pm$1\nl
&1990&+21&157&2,d& \nodata & \nodata & \nodata& \nodata & \nodata & \nodata \nl
&1971&+24&17&1,p&1993&+23&28$\pm$1&1996&+22&26$\pm$1\smallskip \nl
HD 76004& \nodata & \nodata & \nodata & \nodata &1994&$-$20&31$\pm$2& \nodata & \nodata & \nodata \nl
& \nodata & \nodata & \nodata & \nodata &1994&$-$6&19$\pm$5& \nodata & \nodata & \nodata \nl
& \nodata & \nodata & \nodata & \nodata &1994&+5&24$\pm$4& \nodata & \nodata & \nodata \nl
& \nodata & \nodata & \nodata & \nodata &1994&+10&11$\pm$7& \nodata & \nodata & \nodata \nl
& \nodata & \nodata & \nodata & \nodata &1994&+24&11$\pm$2& \nodata & \nodata & \nodata \nl
& \nodata & \nodata & \nodata & \nodata &1994&+33&34$\pm$2& \nodata & \nodata & \nodata \smallskip \nl
HD 76161& \nodata & \nodata & \nodata & \nodata &1993&$-$8&16$\pm$1& \nodata & \nodata & \nodata \nl
&1971&$-$1&34&1,p &\nodata & \nodata & \nodata & \nodata &\nodata & \nodata \nl
&$<$1971&0&51&4,p&1993&+2&38$\pm$2& \nodata & \nodata & \nodata \smallskip \nl
HD 76534&\nodata&\nodata&\nodata&\nodata&1994&$-$86&7$\pm$2& \nodata & \nodata & \nodata \nl
& \nodata & \nodata & \nodata & \nodata &1994&$-$27&16$\pm$1& \nodata & \nodata & \nodata \nl
& \nodata & \nodata & \nodata & \nodata &1994&$-$17&18$\pm$1& \nodata & \nodata & \nodata \nl
& \nodata & \nodata & \nodata & \nodata &1994&$-$4&13$\pm$2& \nodata & \nodata & \nodata \nl
& \nodata & \nodata & \nodata & \nodata &1994&+8&64$\pm$2& \nodata & \nodata & \nodata \nl
& \nodata & \nodata & \nodata & \nodata &1994&+16&28$\pm$3& \nodata & \nodata & \nodata \nl
& \nodata & \nodata & \nodata & \nodata &1994&+32&13$\pm$1& \nodata & \nodata & \nodata \smallskip \nl
HD 76566&1973&$-$18&172&1,p&1994&$-$20&31$\pm$1& \nodata & \nodata & \nodata \nl
& \nodata & \nodata & \nodata & \nodata &1994&$-$10&15$\pm$1& \nodata & \nodata & \nodata \nl
& \nodata & \nodata & \nodata & \nodata &1994&+6&68$\pm$1& \nodata & \nodata & \nodata \nl
& \nodata & \nodata & \nodata & \nodata &1994&+15&53$\pm$2& \nodata & \nodata & \nodata \smallskip \nl
HD 76838& \nodata & \nodata & \nodata & \nodata & \nodata & \nodata & \nodata &1996&$-$18&24$\pm$1\nl
& \nodata & \nodata & \nodata & \nodata & \nodata & \nodata & \nodata &1996&+9&65$\pm$1\nl
& \nodata & \nodata & \nodata & \nodata & \nodata & \nodata & \nodata &1996&+17&7$\pm$1\nl
& \nodata & \nodata & \nodata & \nodata & \nodata & \nodata & \nodata &1996&+33&7$\pm$1\smallskip \nl
HD 77320& \nodata & \nodata & \nodata & \nodata &1994&$-$20&18$\pm$1& \nodata & \nodata & \nodata \nl
& \nodata & \nodata & \nodata & \nodata&1994&$-$8&3$\pm$1& \nodata & \nodata & \nodata \nl
&1971-77&$-$4&34&1,p& \nodata & \nodata & \nodata & \nodata& \nodata & \nodata \nl
& \nodata & \nodata & \nodata & \nodata &1994&+3&16$\pm$1& \nodata & \nodata & \nodata \nl
& \nodata & \nodata & \nodata & \nodata &1994&+36&4$\pm$1& \nodata & \nodata & \nodata \smallskip \nl
HD 78005& \nodata & \nodata & \nodata & \nodata &1994&$-$24&9$\pm$1& \nodata & \nodata & \nodata \nl
& \nodata & \nodata & \nodata & \nodata &1994&+3&135$\pm$3& \nodata & \nodata & \nodata \nl
& \nodata & \nodata & \nodata & \nodata &1994&+19&26$\pm$2$^\star$& \nodata & \nodata & \nodata \nl
& \nodata & \nodata & \nodata & \nodata &1994&+33&35$\pm$2& \nodata & \nodata & \nodata \smallskip \nl
HD 79275& \nodata & \nodata & \nodata & \nodata &1994&$-$28&30$\pm$1& \nodata & \nodata & \nodata \nl
& \nodata & \nodata & \nodata & \nodata &1994&$-$13&78$\pm$1& \nodata & \nodata & \nodata \nl
& \nodata & \nodata & \nodata & \nodata &1994&+1&91$\pm$1& \nodata & \nodata & \nodata 
\enddata
\tablenotetext{}{$^\star$Broad absorption feature with central velocity corresponding to the star's radial velocity.}
\tablenotetext{}{$^\bullet$Broad absorption feature, possibly interstellar in origin.}
\tablenotetext{}{Ref. 1: Wallerstein, Silk, \& Jenkins 1980, CTIO 1.5m coud\'e spectrograph with dispersion = 4.5 or 9 \AA\ mm$^{-1}$ and 9 or 18 \AA\ mm$^{-1}$ for \ion{Ca}{2} and \ion{Na}{1}, respectively.}
\tablenotetext{}{Ref 2: Wallerstein \& Gilroy, 1992,  CTIO 4m echelle spectrograph with R = 20,000.}
\tablenotetext{}{Ref 3: Jenkins, Wallerstein, \& Silk, 1984,  CTIO 4m echelle spectrograph with dispersion = 2.2 \AA\ mm$^{-1}$ and 3.3 \AA\ mm$^{-1}$ for \ion{Ca}{2} and \ion{Na}{1}, respectively.}
\tablenotetext{}{Ref. 4: Wallerstein \& Silk, 1971, CTIO 60-inch telescope and coud\'e spectrograph with R$_{Ca II}$= 20,000 \& R$_{Na I}$= 15,000.}
\tablenotetext{}{Ref. 5: Thackeray \& Warren, 1972,  Radcliffe coud\'e spectrograph with dispersion = 6.8 \AA\ mm$^{-1}$.}
\tablenotetext{}{Ref. 6: Hobbs et al., 1991, McDonald 2.7m coud\'e spectrograph and ESO CAT 1.4m with R$_{1981-1984}$= 55,000 \& R$_{1986-1988}$= 110,000.  Velocities and equivalent widths cited are an average of nine data sets obtained between 1981 and 1988.}
\tablenotetext{}{Detectors used for observations: d--CCD; p--photographic plate; m--multiple observations using two or more of the following detectors: photo-multiplier tube, digicon, reticon, CCD.}

\end{deluxetable}